\documentclass[useAMS,usenatbib]{mn2e}  
\pdfoutput=1
\usepackage{aas_macros}
\usepackage{graphics}
\usepackage{times}
\usepackage{amsmath,amssymb}
\usepackage{bm}

\def\be{\begin{equation}}
\def\ee{\end{equation}}
\def\ba{\begin{eqnarray}}
\def\ea{\end{eqnarray}}

\def\lsim{\mathrel{\rlap{\lower4pt\hbox{\hskip1pt$\sim$}}
    \raise1pt\hbox{$<$}}}                
\def\gsim{\mathrel{\rlap{\lower4pt\hbox{\hskip1pt$\sim$}}
    \raise1pt\hbox{$>$}}}

\pagerange{\pageref{firstpage}--\pageref{lastpage}}
\pubyear{2014}

\begin{document}

\label{firstpage}

\title[Weak lensing using only galaxy position angles]{Weak lensing
  using only galaxy position angles} 
\author[Lee Whittaker, Michael L. Brown \& Richard A. Battye]{Lee
  Whittaker, Michael L. Brown, \& Richard A. Battye\\ 
  Jodrell Bank Centre for Astrophysics, School of Physics and
  Astronomy, University of Manchester, Oxford Road,
  Manchester M13 9PL} \date{\today}

\maketitle

\begin{abstract}
  We develop a method for performing a weak lensing analysis using
  only measurements of galaxy position angles. By analysing the
  statistical properties of the galaxy orientations given a known
  intrinsic ellipticity distribution, we show that it is possible to
  obtain estimates of the shear by minimizing a $\chi^2$
  statistic. The method is demonstrated using simulations where the
  components of the intrinsic ellipticity are taken to be Gaussian
  distributed. Uncertainties in the position angle measurements
  introduce a bias into the shear estimates which can be reduced to
  negligible levels by introducing a correction term into the
  formalism. We generalize our approach by developing an algorithm to
  obtain direct shear estimators given any azimuthally symmetric
  intrinsic ellipticity distribution. We introduce a method of 
  measuring the position angles of the galaxies from noisy 
  pixelized images, and propose a method to correct for biases 
  which arise due to pixelization and correlations between 
  measurement errors and galaxy ellipticities. We also develop 
  a method to constrain the sample of galaxies used to obtain an estimate 
  of the intrinsic ellipticity distribution such that fractional biases 
  in the resulting shear estimates are below a given threshold value. 
  We demonstrate the angle only method by applying it to simulations 
  where the ellipticities are taken to follow a log-normal 
  distribution. We compare the performance of the position angle 
  only method with the standard method based on full ellipticity 
  measurements by reconstructing lensing convergence maps from 
  both numerical simulations and from the CFHTLenS data. We find that the 
  difference between the convergence maps reconstructed using the two methods 
  is consistent with noise. 
\end{abstract}

\begin{keywords}
  methods: statistical - methods: analytical - cosmology: theory -
  weak gravitational lensing
\end{keywords}

\section{Introduction}
\label{sec:intro}
Gravitational lensing is a phenomenon that is predicted by General
Relativity and is a consequence of the curvature of spacetime. Light
rays from distant background objects are deflected by a massive
foreground object, such as a galaxy or a clump of dark matter. This
deflection can be used to estimate the projected mass distribution of
the foreground object. Strong gravitational lensing is concerned with
the large deflection of light from a background galaxy by a massive
foreground object close to the line of sight, which leads to multiple
images of the source galaxy; a phenomenon first observed by
\cite{walsh79}. Weak lensing is concerned with observations where the
deflections of the light rays are much smaller, resulting in small
distortions in the observed shapes of background
galaxies. Cosmological weak lensing (or cosmic shear) aims to detect
the coherent shape distortions in the images of background galaxies due to
the intervening large scale structure of the Universe. This is
achieved by performing a statistical analysis of the observed shapes of
the background galaxies in order to extract noisy estimates of the
weak lensing distortion (or ``shear'') field. The effect is difficult
to detect due to the intrinsic randomness of galaxy shapes, and it was
only conclusively detected at the turn of the Millennium
(\citealt{bacon00, kaiser00, wittman00, vanwaerbeke00}). Since then,
much progress has been made in the precision of the measurements of
galaxy shapes and weak lensing is now established as a powerful
cosmological tool. It has already been used to constrain of a number
of cosmological parameters, such as the amplitude of the matter power
spectrum (e.g. \citealt{brown03, hoekstra06, fu08}) and the dark
energy equation of state (e.g. \citealt{schrabback10, kilbinger13}),
while future surveys will provide unprecedented sensitivity to dark
energy parameters (e.g. \citealt{albrecht06, peacock06}).

The standard method of performing a weak lensing analysis requires
measurements of the ellipticities of a set of background
galaxies. These measurements require the application of complex
correction and/or fitting algorithms (e.g. \citealt{kaiser95,
  bridle02, miller07, kitching08}), which can introduce systematic
biases into the measurements if the point spread function is not
accurately accounted for, or if the prior galaxy model is
incorrect. In order to achieve unbiased ellipticity estimates, these
algorithms generally require the application of additive and
multiplicative calibration corrections derived from simulations (see
e.g. \citealt{heymans12}). If the multiplicative bias is identical for
both components of the ellipticity, then this bias will be absent from
the unit vectors that describe the galaxy orientation. It is
conceivable therefore that measurements of the orientations of
galaxies will not be subject to the multiplicative biases inherent in
the full ellipticity analysis and may consequently be more robust to
residual biases resulting from an incorrect calibration.

This paper describes a method for performing weak lensing using only
the measurements of the position angles of a set of background
galaxies. Based on an original suggestion by \cite{kochanek90}, this
approach was first explored in \cite{schneider95}, where it was
assumed that the modulus of the intrinsic ellipticities follows a
Gaussian distribution. Under this assumption it was shown that the
mean unit vectors describing the galaxy position angles can be written
as a function of the complex distortion. By inverting this
relationship, \cite{schneider95} were able to obtain an estimate of
both the modulus and the orientation of the lensing distortion field.

Working in the regime of weak lensing, this paper develops the ideas
presented in \cite{schneider95}. Under the assumption of an
azimuthally symmetric (in the \{$\epsilon_1, \epsilon_2$\} plane)
intrinsic ellipticity distribution, and a prior knowledge of the
ellipticity dispersion, we develop a $\chi^2$ statistic in Section
\ref{sec:method}, which can be minimized numerically in order to
obtain estimates of the shear. It is found that inherent biases arise
from measurement errors on the position angles. However, a method for
reducing these biases to negligible levels is then proposed. In Section 
\ref{sec:pos_angles} we develop a method of measuring the position angles of 
galaxies from noisy pixelized images. We use the position angle measurements 
to recover shear estimates and compare the performance of this method with the
KSB method, where full ellipticity information is used. In Section \ref{sec:errors_pofe} 
we investigate the impact of an imperfect knowledge of the intrinsic ellipticity 
distribution. We place constraints on the size of the sample of galaxies and the 
errors on the ellipticity measurements used to estimate the distribution necessary 
to ensure that biases in the shear estimates resulting from an imperfect distribution 
are below a given threshold value. We compare the performance of the position angle only
approach with the standard (full ellipticity) approach by performing
mass reconstructions using simulated data (Section \ref{sec:sims}) and
using the data from the Canada France Hawaii Lensing Survey (CFHTLenS,
Section~\ref{sec:CFHT}). We conclude with a discussion in Section
\ref{sec:conclusions}.

\section{Constructing Angle Only Shear Estimators}
\label{sec:method} 
The standard method for performing a weak lensing measurement involves
averaging over the observed ellipticities of a set of galaxies. We
begin by pixelizing the sky, such that we concentrate on an area small
enough that the shear can be considered constant. Working within
the regime of weak lensing we can then express the observed (complex)
ellipticity of a galaxy, 
$\bm{\epsilon}^{\mathrm{obs}} = \epsilon^{\rm obs}_1 + i \, \epsilon^{\rm obs}_2$, 
in terms of the intrinsic
ellipticity of the galaxy, $\bm{\epsilon}^{\mathrm{int}}$, and the
constant reduced shear signal in a given pixel, $\bm{g}$, such that (see
\citealt{bartelmann01} for a discussion)
\begin{equation}\label{eq:obs_ellip_shear}
\bm{\epsilon}^{\mathrm{obs}}=\frac{\bm{\epsilon}^{\mathrm{int}}+\bm{g}}{1+\bm{g}^*\bm{\epsilon}^{\mathrm{int}}},
\end{equation}
If we now assume that the expectation value of the intrinsic
ellipticities, $\left<\bm{\epsilon}^{\mathrm{int}}\right>$, is zero we
can write the standard shear estimator as
\begin{equation}\label{eq:standard_est}
\hat{\bm{g}}=\frac{\sum_{i=1}^Nw_i\bm{\epsilon}_i^{\mathrm{obs}}}{\sum_{i=1}^Nw_i},
\end{equation}
where $w_i$ is a weight, which could, for example, be dependent on the
intrinsic distribution in the ellipticities and ellipticity
measurement errors. If we make the further assumption that the
measurement error on $\bm{\epsilon}^{\mathrm{obs}}$ is much smaller
than the intrinsic dispersion in galaxy ellipticities,
$\sigma_{\epsilon}$, then uniform weighting ($w_i=1$) is an optimal
choice. In this case, the error on the standard estimator is a result
of the intrinsic shape dispersion only, i.e.,
\begin{equation}\label{eq:standard_est_err}
\sigma_{\hat{\bm{g}}}=\frac{\sigma_{\epsilon}}{\sqrt{N}}.
\end{equation}
Denoting the observed position angle as $\alpha$, we can express the
observed ellipticity in polar form, such that
\begin{align}\label{eq:obs_ellip_comp_angle}
\epsilon_1^{\mathrm{obs}}&=\left|\bm{\epsilon}^{\mathrm{obs}}\right|\cos\left(2\alpha\right),\nonumber\\
\epsilon_2^{\mathrm{obs}}&=\left|\bm{\epsilon}^{\mathrm{obs}}\right|\sin\left(2\alpha\right).
\end{align}
Let us assume that the distribution of the intrinsic ellipticities of
the galaxies can be described by an azimuthally symmetric probability
density function,
$f\left(\left|\bm{\epsilon}^{\mathrm{int}}\right|\right)$. As the
shear in a pixel is constant, this implies that the observed
ellipticity can be modelled as
\begin{equation}\label{eq:ia.1}
\bm{\epsilon}^{\mathrm{obs}}=\bm{g}+\bm{\epsilon}^{\mathrm{ran}},
\end{equation}
where $\bm{\epsilon}^{\mathrm{ran}}$ is a random vector which is dependent on the
intrinsic distribution, with a mean of zero.  In this paper we are
interested in using measurements of the galaxy position angles
($\alpha$) alone to estimate the shear. We must, therefore, consider
the statistics of the sine and cosine functions. Defining the
components of the shear as
\begin{align}\label{eq:ia_comp}
g_1&=\left|\bm{g}\right|\cos\left(2\alpha_0\right),\nonumber\\
g_2&=\left|\bm{g}\right|\sin\left(2\alpha_0\right),
\end{align} 
it can be shown that, for any distribution of $\bm{\epsilon}^{\mathrm{ran}}$ 
which exhibits reflection symmetry about the vector $\bm{g}$\footnote{$\bm{\epsilon}^{\mathrm{ran}}$ is symmetrically distributed about $\bm{g}$ for any azimuthally symmetric intrinsic ellipticity distribution.}, 
the mean of the cosines and sines of the position
angles, $\left<\cos\left(2\alpha\right)\right>$ and
$\left<\sin\left(2\alpha\right)\right>$, can be written as
\begin{align}\label{eq:mean_gen_trig}
\left<\cos\left(2\alpha\right)\right>=&F_1\left(\left|\bm{g}\right|\right)\cos\left(2\alpha_0\right),\nonumber\\
\left<\sin\left(2\alpha\right)\right>=&F_1\left(\left|\bm{g}\right|\right)\sin\left(2\alpha_0\right).
\end{align}
Defining $\left|\bm{\epsilon}_{\mathrm{max}}^{\mathrm{int}}\right|$ as
the maximum value of the modulus of the intrinsic ellipticity, the
function $F_1\left(\left|\bm{g}\right|\right)$ can be written in
terms of the intrinsic ellipticity distribution, such that
\begin{align}\label{eq:general_F}
F_1\left(\left|\bm{g}\right|\right)=\, &\frac{1}{\pi}\int_0^{\left|\bm{\epsilon}_{\mathrm{max}}^{\mathrm{int}}\right|}\int_{-\frac{\pi}{2}}^{\frac{\pi}{2}}\mathrm{d}\alpha^{\mathrm{int}}\mathrm{d}\left|\bm{\epsilon}^{\mathrm{int}}\right|f\left(\left|\bm{\epsilon}^{\mathrm{int}}\right|\right)\nonumber\\
&\times h_1\left(\left|\bm{g}\right|,\left|\bm{\epsilon}^{\mathrm{int}}\right|,\alpha^{\mathrm{int}}\right),
\end{align}
where $\alpha^{\rm int}$ is the intrinsic position angle and the
function
$h_1\left(\left|\bm{g}\right|,\left|\bm{\epsilon}^{\mathrm{int}}\right|,\alpha^{\mathrm{int}}\right)$
is found to be
\begin{equation}\label{eq:g_function}
h_1\left(\left|\bm{g}\right|,\left|\bm{\epsilon}^{\mathrm{int}}\right|,\alpha^{\mathrm{int}}\right)=
\frac{\epsilon_1'}{\sqrt{\epsilon_1'^2+\epsilon_2'^2}},
\end{equation}
with
\begin{align}\label{eq:e'}
\epsilon_1'=&\left|\bm{g}\right|\left(1+\left|\bm{\epsilon}^{\mathrm{int}}\right|^2\right)+\left(1+\left|\bm{g}\right|^2\right)\left|\bm{\epsilon}^{\mathrm{int}}\right|\cos\left(2\alpha^{\mathrm{int}}\right)\nonumber\\
\epsilon_2'=&\left(1-\left|\bm{g}\right|^2\right)\left|\bm{\epsilon}^{\mathrm{int}}\right|\sin\left(2\alpha^{\mathrm{int}}\right)
\end{align}
 From equation (\ref{eq:mean_gen_trig}) it is clear that we can
 estimate the orientation of the shear as
\begin{equation}\label{eq:alpha_0}
\alpha_0=\frac{1}{2}\tan^{-1}\left(\frac{\left<\sin\left(2\alpha\right)\right>}{\left<\cos\left(2\alpha\right)\right>}\right),
\end{equation}
which is equal to the mean observed position angle, $\left<\alpha\right>$.

Letting $2\alpha=\theta$ and $2\alpha_0=\theta_0$, a more general form
of equation~(\ref{eq:mean_gen_trig}) can be written as
\begin{align}\label{eq:gen_mean_trig_Fn}
\left<\cos\left(n\theta\right)\right>&=F_n\left(\left|\bm{g}\right|\right)\cos\left(n\theta_0\right),\nonumber\\
\left<\sin\left(n\theta\right)\right>&=F_n\left(\left|\bm{g}\right|\right)\sin\left(n\theta_0\right),
\end{align}
where $n$ is any positive integer.

By considering a general function
$h_n\left(\left|\bm{g}\right|,\left|\bm{\epsilon}^{\mathrm{int}}\right|,\alpha^{\mathrm{int}}\right)$,
we can write the general $F_n\left(\left|\bm{g}\right|\right)$
function for any azimuthally symmetric intrinsic probability
distribution as
\begin{align}\label{eq:gen_Fn}
F_n\left(\left|\bm{g}\right|\right)=\, &\frac{1}{\pi}\int_0^{\left|\bm{\epsilon}_{\mathrm{max}}^{\mathrm{int}}\right|}\int_{-\frac{\pi}{2}}^{\frac{\pi}{2}}\mathrm{d}\alpha^{\mathrm{int}}\mathrm{d}\left|\bm{\epsilon}^{\mathrm{int}}\right|f\left(\left|\bm{\epsilon}^{\mathrm{int}}\right|\right)\nonumber\\
&\times h_n\left(\left|\bm{g}\right|,\left|\bm{\epsilon}^{\mathrm{int}}\right|,\alpha^{\mathrm{int}}\right).
\end{align}
The $F_2\left(\left|\bm{g}\right|\right)$ function will be useful
in later discussions regarding the variance of
$\cos\left(2\alpha\right)$ and $\sin\left(2\alpha\right)$. The
corresponding
$h_2\equiv h_2\left(\left|\bm{g}\right|,\left|\bm{\epsilon}^{\mathrm{int}}\right|,\alpha^{\mathrm{int}}\right)$
function is
\begin{equation}
h_2=\frac{\epsilon_1'^2-\epsilon_2'^2}{\epsilon_1'^2+\epsilon_2'^2}
\end{equation}
From equation (\ref{eq:mean_gen_trig}) it is clear that the mean sine
and cosine functions trace the sine and cosine of the shear
orientation, subject to a scaling factor which depends on
$\left|\bm{g}\right|$.

\subsection{Shear estimation using $\chi^2$ minimization}
\label{subsec:chi2}
It is possible to obtain constraints on the shear parameters from
measurements of the galaxy position angles, $\alpha^{(i)}$, alone using the
least-squares method \citep{Press.1992}. For any azimuthally symmetric
intrinsic ellipticity distribution, equation~(\ref{eq:mean_gen_trig})
suggests the definition of a general $\chi^2$ as
\begin{equation}\label{eq:initialchi^2}
\chi^2=\sum_{i,j=1}^N\left(\bm{n}^{(i)}-F_1\left(\left|\bm{g}\right|\right)\frac{\bm{g}}{\left|\bm{g}\right|}\right)^{\mathrm{T}}\mathbf{C}_{ij}\left(\bm{n}^{(j)}-F_1\left(\left|\bm{g}\right|\right)\frac{\bm{g}}{\left|\bm{g}\right|}\right),
\end{equation}
where we have defined the observed unit vector for the $i^{\mathrm{th}}$ galaxy as
\begin{equation}\label{eq:unit_vector}
\bm{n}^{(i)}=\left(
\begin{array}{c}
\cos(2\alpha^{(i)})\\
\sin(2\alpha^{(i)})
\end{array}\right),
\end{equation}
and $\mathbf{C}$ is the covariance matrix. 

For the case when there are no measurement
errors on $\alpha$, it can be shown, using equation
(\ref{eq:gen_mean_trig_Fn}), that the variance on the unit vector
components is
\begin{equation}\label{eq:variance_F}
\sigma_{n_{1,2}}^2=\frac{1}{2}\left(1-F_1^2\right)\pm\frac{1}{2}\left(F_2-F_1^2\right)\cos\left(4\alpha_0\right),
\end{equation}
where we define $F_k\equiv F_k\left(\left|\bm{g}\right|\right)$,
and the plus and minus signs correspond to the first and second
components of $\bm{n}$ respectively. In the limit of zero shear the
value of the variance is 0.5 for both components as there is no
preferred observed position angle.

The form of the covariance is found to be
\begin{equation}\label{eq:cov_F}
\mathrm{cov}\left(\cos\left(2\alpha\right),\sin\left(2\alpha\right)\right)=\frac{1}{2}\left(F_2-F_1^2\right)\sin\left(4\alpha_0\right),
\end{equation}
which is zero in the limit of zero shear. 

\begin{figure}
\centering
\includegraphics{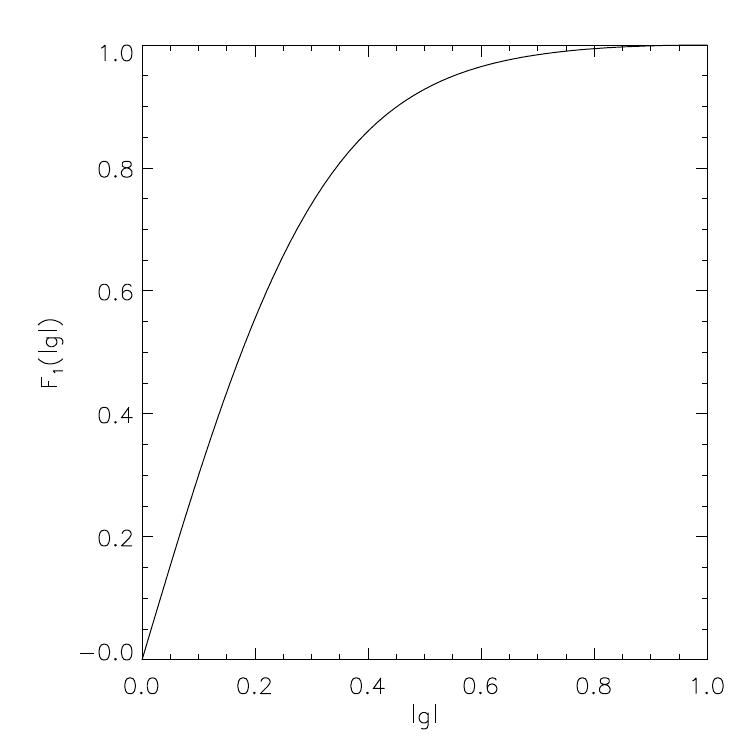}
\caption{The $F_1\left(\left|\bm{g}\right|\right)$ function for case of the Rayleigh distribution given in equation (\ref{eq:gauss_dist}).}
\label{fig:Ffunc_gauss}
\end{figure}
If the components of the intrinsic ellipticity are Gaussian
distributed, then $\left|\bm{\epsilon}^{\mathrm{int}}\right|$ is
Rayleigh distributed (for a discussion on the motivation for this
form of distribution see, for example, \citealt{viola13}):
\begin{equation}\label{eq:gauss_dist}
f\left(\left|\bm{\epsilon}^{\mathrm{int}}\right|\right)=\frac{\left|\bm{\epsilon}^{\mathrm{int}}\right|}{\sigma_{\epsilon}^2\left(1-\exp\left(-\frac{\left|\bm{\epsilon}_{\mathrm{max}}^{\mathrm{int}}\right|^2}{2\sigma_{\epsilon}^2}\right)\right)}\exp\left(-\frac{\left|\bm{\epsilon}^{\mathrm{int}}\right|^2}{2\sigma_{\epsilon}^2}\right).
\end{equation}
Using this form for the distribution of $\left|\bm{\epsilon}^{\mathrm{int}}\right|$, with $\sigma_{\epsilon}=0.3/\sqrt{2}$ 
and $\left|\bm{\epsilon}_{\mathrm{max}}^{\mathrm{int}}\right|=1.0$, we constructed $F_1\left(\left|\bm{g}\right|\right)$, shown in Figure \ref{fig:Ffunc_gauss}, and $F_2\left(\left|\bm{g}\right|\right)$. Using these two functions we found that, for shear values in the range $\left|\bm{g}\right|\leq0.1$, equation (\ref{eq:variance_F})
gives a variance in the range $0.44\lesssim\sigma_{\bm{n}}^2\leq0.5$,
while equation (\ref{eq:cov_F}) predicts a covariance in the range
$-0.014\lesssim\mathrm{cov}\left(\cos\left(2\alpha\right),\sin\left(2\alpha\right)\right)\lesssim0.014$. For
the subsequent numerical calculations we use
$\sigma^2=\sigma_{\bm{n}}^2=0.5$ and
$\mathrm{cov}\left(\cos\left(2\alpha\right),\sin\left(2\alpha\right)\right)=0$
in every $\chi^2$ that we construct. If we make the further assumption
that the measurements of the position angles of different galaxies are
independent we can simplify the $\chi^2$, such that it now takes the
form
\begin{equation}\label{eq:simp_F_chi}
\chi^2=\sum_{i=1}^N\frac{\left|\bm{n}^{(i)}-F_1\left(\left|\bm{g}\right|\right)\frac{\bm{g}}{\left|\bm{g}\right|}\right|^2}{\sigma^2}.
\end{equation}
The minimization of this $\chi^2$ gives us an estimate of the shear,
$\hat{\bm{g}}$, which satisfies the equations
\begin{align}\label{eq:est_cond_full}
\frac{1}{N}\sum_{i=1}^N\cos(2\alpha^{(i)})&=F_1\left(\left|\hat{\bm{g}}\right|\right)\frac{\hat{g}_1}{\left|\hat{\bm{g}}\right|},\nonumber\\
\frac{1}{N}\sum_{i=1}^N\sin(2\alpha^{(i)})&=F_1\left(\left|\hat{\bm{g}}\right|\right)\frac{\hat{g}_2}{\left|\hat{\bm{g}}\right|}.
\end{align}
Taking the ratio of these two equations yields an estimate of the orientation of the shear:
\begin{equation}\label{eq:orient_shear}
\frac{\hat{g}_2}{\hat{g}_1} =
\frac{\sum_{i=1}^N\sin(2\alpha^{(i)})}
     {\sum_{i=1}^N\cos(2\alpha^{(i)})}.
\end{equation}
If, instead, we square and sum them together we obtain an estimate of
the $F_1\left(\left|\bm{g}\right|\right)$ function which, in
turn, depends on the modulus of the shear:
\begin{equation}\label{eq:mod_shear}
F_1\left(\left|\hat{\bm{g}}\right|\right)=\sqrt{\left[\frac{1}{N}\sum_{i=1}^N\cos\left(2\alpha^{(i)}\right)\right]^2+\left[\frac{1}{N}\sum_{i=1}^N\sin\left(2\alpha^{(i)}\right)\right]^2}.
\end{equation}
Equation (\ref{eq:mod_shear}) can be solved numerically to yield an
estimate of the shear modulus using, for example, the Secant method
\citep{Press.1992}, or by tabulating $F_1\left(\left|\bm{g}\right|\right)$ and inverting the function to recover $\left|\hat{\bm{g}}\right|$. By combining equations (\ref{eq:orient_shear}) and
(\ref{eq:mod_shear}) we can therefore obtain estimates for
both components of the shear. The error on the shear estimates can then be estimated as
\begin{equation}\label{eq:theor_variance}
\sigma_{\hat{\bm{g}}}=\frac{\left|\hat{\bm{g}}\right|}{F_1\left(\left|\hat{\bm{g}}\right|\right)}\frac{\hat{\sigma}_{\bm{n}}}{\sqrt{N}},
\end{equation}
where $\hat{\sigma}_{\bm{n}}$ is found by substituting the
estimated shear values into equation (\ref{eq:variance_F}).
\begin{figure*}
\begin{minipage}{6in}
\centering
\includegraphics{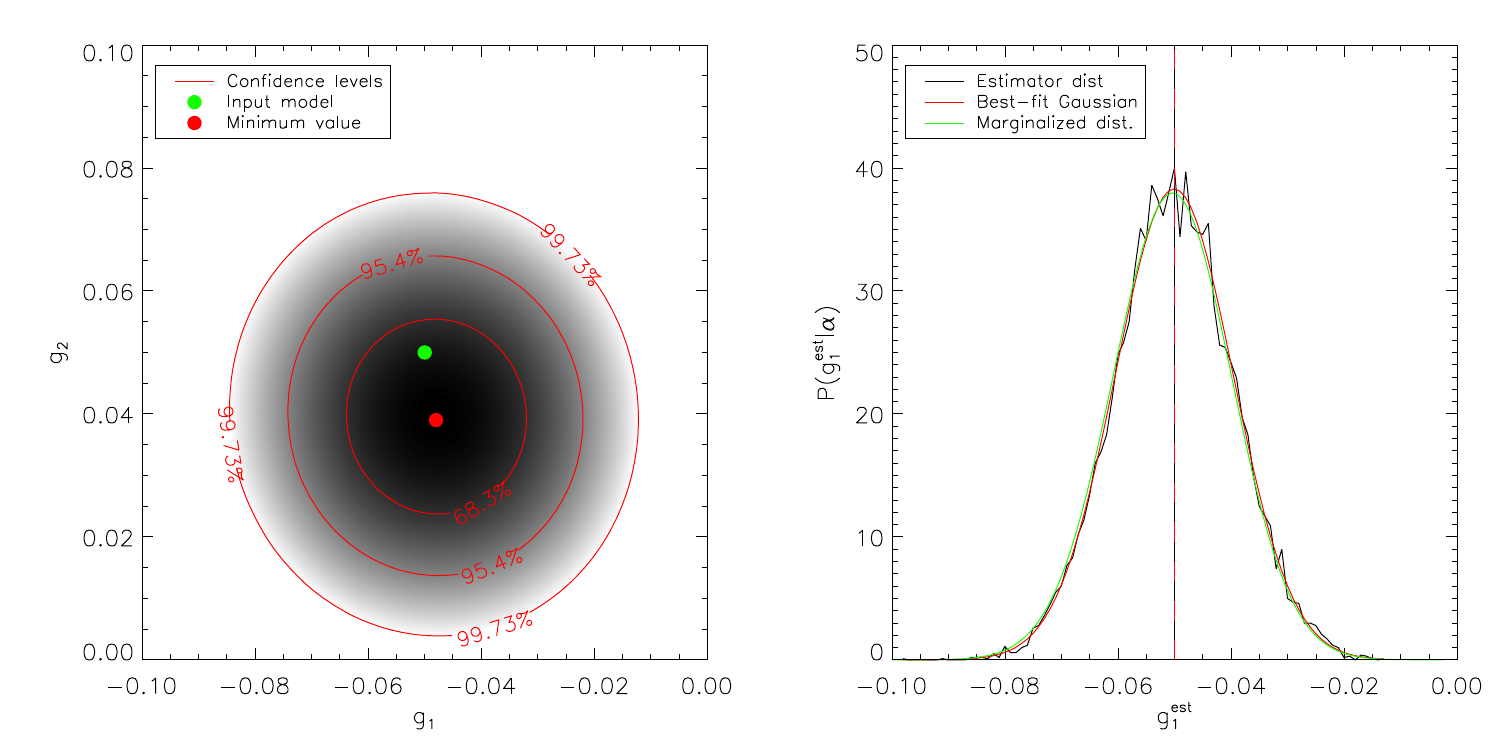}
\caption{\emph{Left panel:} The grey-scale shows the $\chi^2$ of
  equation (\ref{eq:simp_F_chi}) recovered from a set of simulated
  galaxy position angles with input shear values of $g_1=-0.05$
  and $g_2=0.05$ and where we assume a Rayleigh distribution for
  $\left|\bm{\epsilon}^{\mathrm{int}}\right|$. The best-fit shear is
  given by the minimum value of the $\chi^2$ and the contours show the
  68.3\%, 95.4\% and 99.73\% confidence levels, plotted under the
  assumption that the shear estimates are Gaussian
  distributed. \emph{Right panel:} The distribution of the best-fit
  $\hat{g}_1$ values obtained from $10^4$ realizations. The red
  curve is a Gaussian distribution constructed using the variance and
  the mean of the best-fit values. The green curve shows a
  marginalized plot of the distribution of the left panel, and hence
  is the distribution of $g_1$ for one realization. The agreement
  between the curves demonstrates that the shear estimates are approximately 
  Gaussian distributed and validates the use of the $\chi^2$ contours
  plotted in the left panel.}
\label{fig:likelihood_no_sig}
\end{minipage}
\end{figure*}

Using a simulation composed of 500 galaxies with input shear values of
$g_1=-0.05$ and $g_2=0.05$, and assuming a Rayleigh
distribution for $\left|\bm{\epsilon}^{\mathrm{int}}\right|$ with
$\sigma_{\epsilon}=0.3/\sqrt{2}$, we tested the performance of this
method. A zero measurement error on $\alpha$ was assumed for this
initial test. To estimate the shear, we performed a grid-based search
over the shear parameters, calculating the $\chi^2$ corresponding
to each parameter value. The best-fit values were those which
minimized the $\chi^2$ statistic and these were found to be consistent
with equation (\ref{eq:est_cond_full}). The results of the test are
shown in the left panel of Fig.~\ref{fig:likelihood_no_sig}. The
best-fit values were found to be $\hat{g}_1=-0.048$ and
$\hat{g}_2=0.039$, with $\chi^2=964$; using 500 galaxies we
expect a value of $\chi^2\approx1000$, therefore this value of
$\chi^2$ is consistent with our model providing a good fit to the
data. The right hand panel of Fig.~\ref{fig:likelihood_no_sig} shows
the distribution of the best-fit $g_1$ values obtained over $10^4$
realizations; with each realization consisting of 500 galaxies and
where we have used a bin size of $\Delta\hat{g}_1=0.001$.
\begin{figure}
\centering
\includegraphics{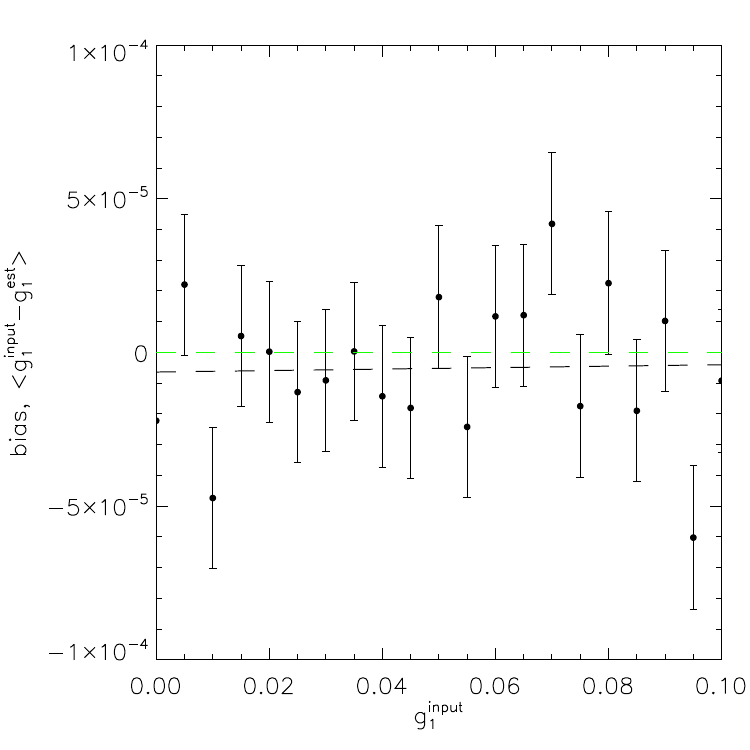}
\caption{Residual bias in the shear estimates obtained using equations
  (\ref{eq:orient_shear}) and (\ref{eq:mod_shear}) as a function of
  the input shear value, $g_1$. For each simulation, the input
  value of $g_2$ was randomly selected from a uniform
  distribution in the range $-0.1\leq g_2\leq0.1$. The green
  dashed line shows the line of zero bias and the black dashed line is
  the line of best fit for the data. The line of best fit is
  consistent with an overall bias of zero.}
\label{fig:likelihood_no_sig_bias}
\end{figure}

We also ran simulations for a range of input shear values in order to
check for biases. For each input shear value, we performed $10^4$
simulations with each simulation composed of $10^4$ galaxies. The
result is shown in Fig.~\ref{fig:likelihood_no_sig_bias}. Note that in
order to obtain the shear estimates for this test, rather than using
the grid-based approach, we used equation (\ref{eq:orient_shear}) to
estimate the shear orientation and we solved equation
(\ref{eq:mod_shear}) by tabulating $F_1\left(\left|\bm{g}\right|\right)$ and then inverting the function to find
the modulus of the shear.\footnote{A small bias is expected in
this approach due to the numerical integration of the
$F_1\left(\left|\bm{g}\right|\right)$ function which is performed
during the estimation process. However, this bias was found to be
negligible in all of the tests that we have performed.}

\subsection{Removal of noise bias}
\label{subsec:rem_noise_chi2}
Estimators based on equation (\ref{eq:simp_F_chi}) are found to be
biased in the presence of measurement errors on $\alpha$. We can write the measured
position angles as $\hat{\alpha}=\alpha+\delta\alpha$, where
$\delta\alpha$ is a random measurement error on the position angle. It is possible to correct for the resulting noise bias by examining the averages of equation
(\ref{eq:est_cond_full}). Allowing for the measurement error on $\alpha$ we have
\begin{align}\label{eq:est_trig_err}
\left<\cos\left(2\hat{\alpha}\right)\right>&=\left<\cos\left(2\alpha+2\delta\alpha\right)\right>,\nonumber\\
\left<\sin\left(2\hat{\alpha}\right)\right>&=\left<\sin\left(2\alpha+2\delta\alpha\right)\right>.
\end{align}
If we take the limit as $N\rightarrow\infty$ and assume that $\delta\alpha$ is independent of
$\alpha$ then we can expand the trigonometric functions, such that
\begin{align}\label{eq:est_trig_bias1}
\left<\cos\left(2\hat{\alpha}\right)\right>&=\left<\cos\left(2\alpha\right)\right>\left<\cos\left(2\delta\alpha\right)\right>-\left<\sin\left(2\alpha\right)\right>\left<\sin\left(2\delta\alpha\right)\right>,\nonumber\\
\left<\sin\left(2\hat{\alpha}\right)\right>&=\left<\sin\left(2\alpha\right)\right>\left<\cos\left(2\delta\alpha\right)\right>+\left<\cos\left(2\alpha\right)\right>\left<\sin\left(2\delta\alpha\right)\right>.
\end{align}
If we further assume that the error distribution is symmetric about zero, then
\begin{align}\label{eq:est_trig_bias1_zeromean}
\left<\cos\left(2\hat{\alpha}\right)\right>&=\left<\cos\left(2\alpha\right)\right>\left<\cos\left(2\delta\alpha\right)\right>,\nonumber\\
\left<\sin\left(2\hat{\alpha}\right)\right>&=\left<\sin\left(2\alpha\right)\right>\left<\cos\left(2\delta\alpha\right)\right>.
\end{align}
Upon defining
\begin{equation}\label{eq:beta_def}
\beta\equiv\left<\cos\left(2\delta\alpha\right)\right>,
\end{equation}
we can invert equation (\ref{eq:est_trig_bias1_zeromean}) and correct for the bias, so that the corrected mean unit vector is now given by
\begin{equation}\label{eq:est_trig_cor}
\left<\bm{n}\right>^{\mathrm{corrected}}=\frac{\left<\bm{n}\right>}{\beta}.
\end{equation}
For the specific case where $\delta\alpha$ is a Gaussian distributed measurement error with zero mean and variance $\sigma_{\alpha}^2$, it can be shown that
\begin{equation}\label{eq:beta_gauss}
\beta=\exp\left(-2\sigma_{\alpha}^2\right).
\end{equation}

We can incorporate this correction into the formulation of the
$\chi^2$ by defining
\begin{equation}\label{eq:simp_F_chi_corr}
\chi^2=\sum_{i=1}^N\frac{\left|\bm{n}^{(i)}-F_1\left(\left|\bm{g}\right|\right)\frac{\bm{g}}{\left|\bm{g}\right|}\beta\right|^2}{\sigma^2}.
\end{equation}
Equation (\ref{eq:simp_F_chi_corr}) is minimized when
\begin{align}\label{eq:est_cond_corr}
\frac{1}{N}\sum_{i=1}^N\cos\left(2\hat{\alpha}^{(i)}\right)&=F_1\left(\left|\hat{\bm{g}}\right|\right)\frac{\hat{g}_1}{\left|\hat{\bm{g}}\right|}\beta,\nonumber\\
\frac{1}{N}\sum_{i=1}^N\sin\left(2\hat{\alpha}^{(i)}\right)&=F_1\left(\left|\hat{\bm{g}}\right|\right)\frac{\hat{g}_2}{\left|\hat{\bm{g}}\right|}\beta.
\end{align}
Following the same procedure as for the case of $\sigma_{\alpha}=0$ in
the previous section, we can now estimate the orientation of the shear as
\begin{equation}\label{eq:orient_shear_corr}
\frac{\hat{g}_2}{\hat{g}_1}=\frac{\sum_{i=1}^N\sin(2\hat{\alpha}^{(i)})}{\sum_{i=1}^N\cos(2\hat{\alpha}^{(i)})},
\end{equation}
while the estimate of the $F_1\left(\left|\bm{g}\right|\right)$
function, which depends on the modulus of the shear, becomes
\begin{equation}\label{eq:mod_shear_corr}
F_1\left(\left|\hat{\bm{g}}\right|\right)=\frac{1}{\beta}\sqrt{\left[\frac{1}{N}\sum_{i=1}^N\cos\left(2\hat{\alpha}^{(i)}\right)\right]^2+\left[\frac{1}{N}\sum_{i=1}^N\sin\left(2\hat{\alpha}^{(i)}\right)\right]^2}.
\end{equation}
From equations (\ref{eq:orient_shear_corr}) and
(\ref{eq:mod_shear_corr}) we see that the estimator for the shear
orientation remains unchanged in the presence of a measurement error on the position angle. However, the expression for $F_1\left(\left|\bm{g}\right|\right)$ is modified;
failing to include this noise correction term will result in an
estimate of $\left|\bm{g}\right|$ that is too small.

We can also examine the impact on the variance on the trigonometric
functions when a measurement error on $\alpha$ is
included. This variance is found to be
\begin{equation}\label{eq:variance_F_corr}
\sigma_{n_{1,2}}^2=\frac{1}{2}\left(1-F_1^2\beta^2\right)\pm\frac{1}{2}\left(F_2\beta_2-F_1^2\beta^2\right)\cos\left(4\alpha_0\right),
\end{equation}
where the plus and minus signs correspond to the first and second
components of $\bm{n}$ respectively and where we have defined
\begin{equation}\label{eq:def_beta2}
\beta_2\equiv\left<\cos\left(4\delta\alpha\right)\right>.
\end{equation}
For the case of a Gaussian measurement error on the position angle $\beta_2=\exp\left(-8\sigma_{\alpha}^2\right)$. By substituting the estimated
shear values into equation (\ref{eq:variance_F_corr}), we can estimate
the error on the corrected shear estimator as
\begin{equation}\label{eq:theor_variance_corr}
\sigma_{\hat{\bm{g}}}=\frac{\left|\hat{\bm{g}}\right|}{F_1\left(\left|\hat{\bm{g}}\right|\right)\beta}\frac{\hat{\sigma}_{\bm{n}}}{\sqrt{N}}.
\end{equation}
\begin{table}
\centering
\begin{tabular}{|c|c|c|c|}
\hline
Estimator & $\sigma_{\hat{g}_1}$ & $\sigma_{\hat{g}_1}^{\mathrm{theory}}$ & $\left<\hat{g}_1\right>$ \\[0.5ex]
\hline
original est. & 0.0103 & 0.0102 & $-0.0436\pm0.0001$ \\
corrected est & 0.0119 & 0.0117 & $-0.0502 \pm0.0001$ \\ [1ex]
\hline
\end{tabular}
\caption{The mean and standard deviation of the shear estimates
  recovered from $10^4$ simulations. Values are quoted for both the 
  original $\chi^2$ (equation (\ref{eq:simp_F_chi})) and the corrected
  $\chi^2$ (equation (\ref{eq:simp_F_chi_corr})). The input shear
  value used was $g_1=-0.05$.}
\label{table:10000res}
\end{table}
\begin{figure*}
\begin{minipage}{6in}
\centering
\includegraphics{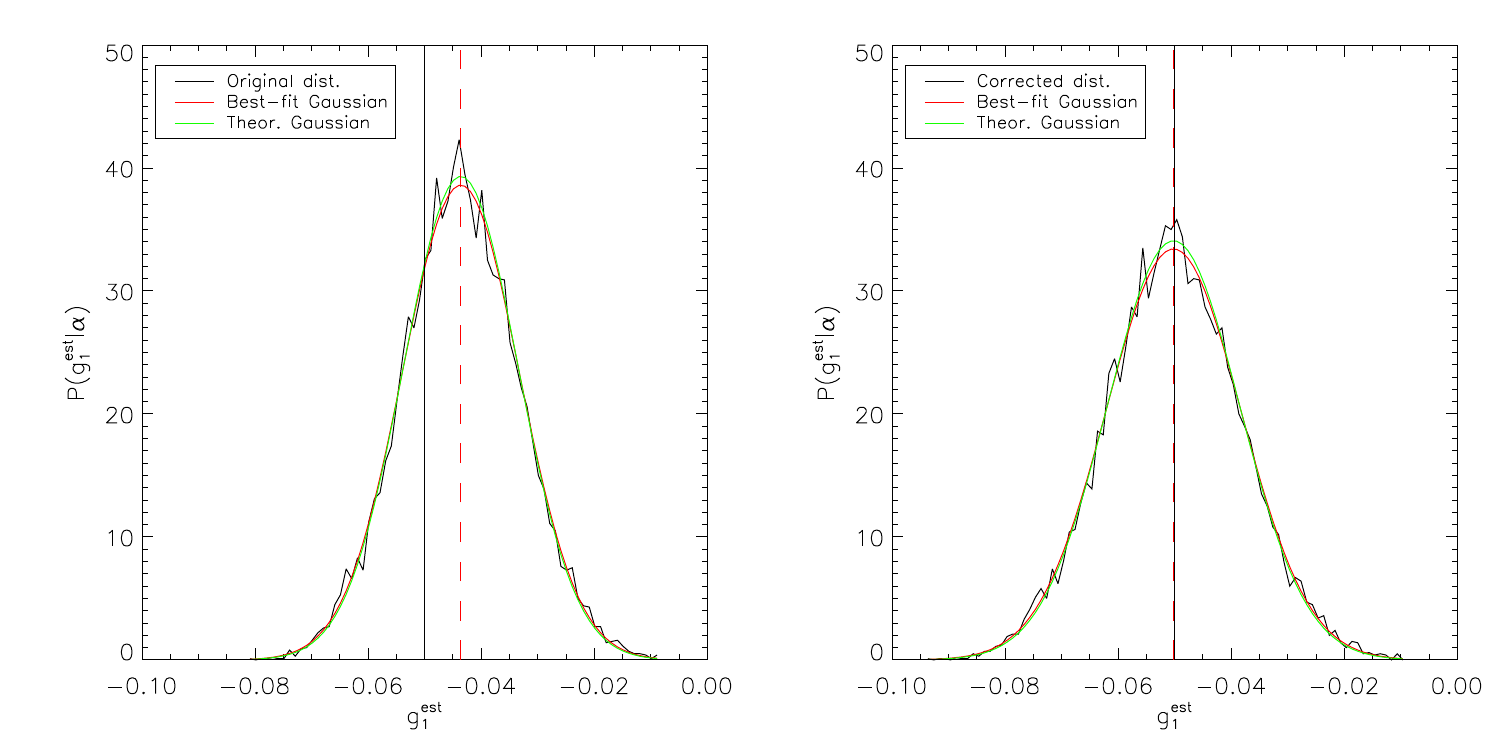}
\caption{The distribution of the best-fit estimates obtained using the
  original $\chi^2$ of equation (\ref{eq:simp_F_chi}) (\emph{left
    panel}) and the corrected form of the $\chi^2$ of equation
  (\ref{eq:simp_F_chi_corr}) (\emph{right panel}) in the presence of a
  Gaussian-distributed measurement error on the galaxy position angles
  with $\sigma_{\alpha}=15^{\circ}$. The simulations consisted of
  $10^4$ realizations, with 500 galaxies in each realization, and assumed a Rayleigh distribution for
  $\left|\bm{\epsilon}^{\mathrm{int}}\right|$. The vertical black line
  shows the input shear value and the red dashed line shows the mean
  recovered best-fit value. The red curves are Gaussian distributions
  with the mean and variance of the estimators. The green curves are
  Gaussian distributions using the input shear value to obtain
  theoretical predictions for the variance from equations
  (\ref{eq:theor_variance}) and (\ref{eq:theor_variance_corr}) for the
  left and right panels respectively. This figure demonstrates that
  the bias introduced by measurement errors on the position angles is
  reduced to negligible levels when the corrected form of the $\chi^2$
  given in equation (\ref{eq:simp_F_chi_corr}) is used. It also
  indicates that equations (\ref{eq:theor_variance}) and
  (\ref{eq:theor_variance_corr}) provide good descriptions of the
  errors in both cases.}
\label{fig:comp_plots_likelihood}
\end{minipage}
\end{figure*}
\begin{figure*}
\begin{minipage}{6in}
\centering
\includegraphics{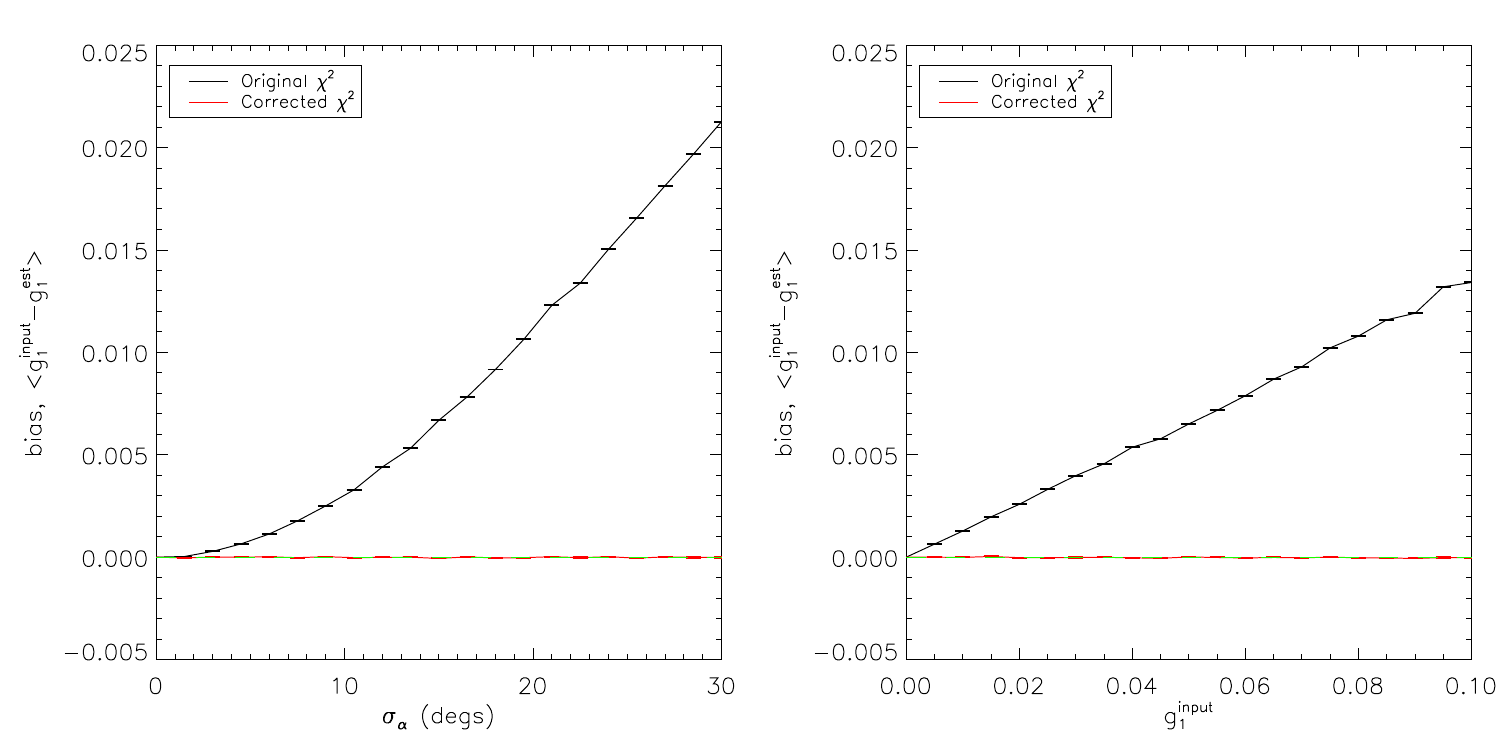}
\caption{The residual bias in the best-fit shear estimates in the presence 
  of a Gaussian measurement error on $\alpha$ and obtained by
  minimizing the two forms of the $\chi^2$ (equations
  (\ref{eq:simp_F_chi}) and (\ref{eq:simp_F_chi_corr})). For each
  point plotted we used $10^4$ realizations, with each realization
  composed of $10^4$ galaxies, in order to suppress numerical error. The left panel shows
  the bias as a function of $\sigma_{\alpha}$ with $g_1=0.05$ and
  with $g_2$ uniformly distributed in the range
  $-0.1\leq g_2\leq0.1$. The right panel shows the bias as a
  function of $g_1$ with $\sigma_{\alpha}=15^{\circ}$ and where
  $g_2$ is again uniformly distributed in the range
  $-0.1\leq g_2\leq0.1$ These figures show that the noise bias due
  to measurement errors on the position angles is reduced
  to negligible levels when the bias correction is used.}
\label{fig:shear_bias_likelihood}
\end{minipage}
\end{figure*}

Fig.~\ref{fig:comp_plots_likelihood} demonstrates the reduction in
bias when this correction is applied to simulations that include
measurement errors on the galaxy position angles. Table \ref{table:10000res} shows
the mean estimated value of $g_1$, $\left<\hat{g}_1\right>$,
and the standard deviation in the estimated values,
$\sigma_{\hat{g}_1}$, obtained from the $10^4$ realizations that
were used to produce Fig.~\ref{fig:comp_plots_likelihood}. The right
column shows how using the corrected form of the $\chi^2$ greatly
reduces the bias introduced by the $15^{\circ}$ error on the position
angle measurements. However, from the left column we see that there is
a modest increase ($\sim$16\%) in the dispersion of the estimates. The
middle column shows the theoretical dispersion in the estimators
obtained using the input values with equations
(\ref{eq:theor_variance}) and (\ref{eq:theor_variance_corr}).

We also ran the simulation for a range of input shear values and for a
range of Gaussian measurement error
values. Fig.~\ref{fig:shear_bias_likelihood} shows the residual bias
in the derived shear estimates as a function of these two
quantities. These results show that the bias in the uncorrected
estimator (equation~\ref{eq:simp_F_chi}) increases approximately
linearly with the input shear, and exponentially with the measurement
error. The success of the correction obtained in equation
(\ref{eq:est_trig_cor}) is clearly demonstrated in this figure.

\section{Measuring position angles}
\label{sec:pos_angles}
In this section we introduce a method of estimating the position angles directly from the data. We use this method to recover a constant shear signal from sets of simulated galaxy images and compare these results with those obtained using the KSB method.
\begin{figure*}
\begin{minipage}{6.48in}
\centering
\includegraphics{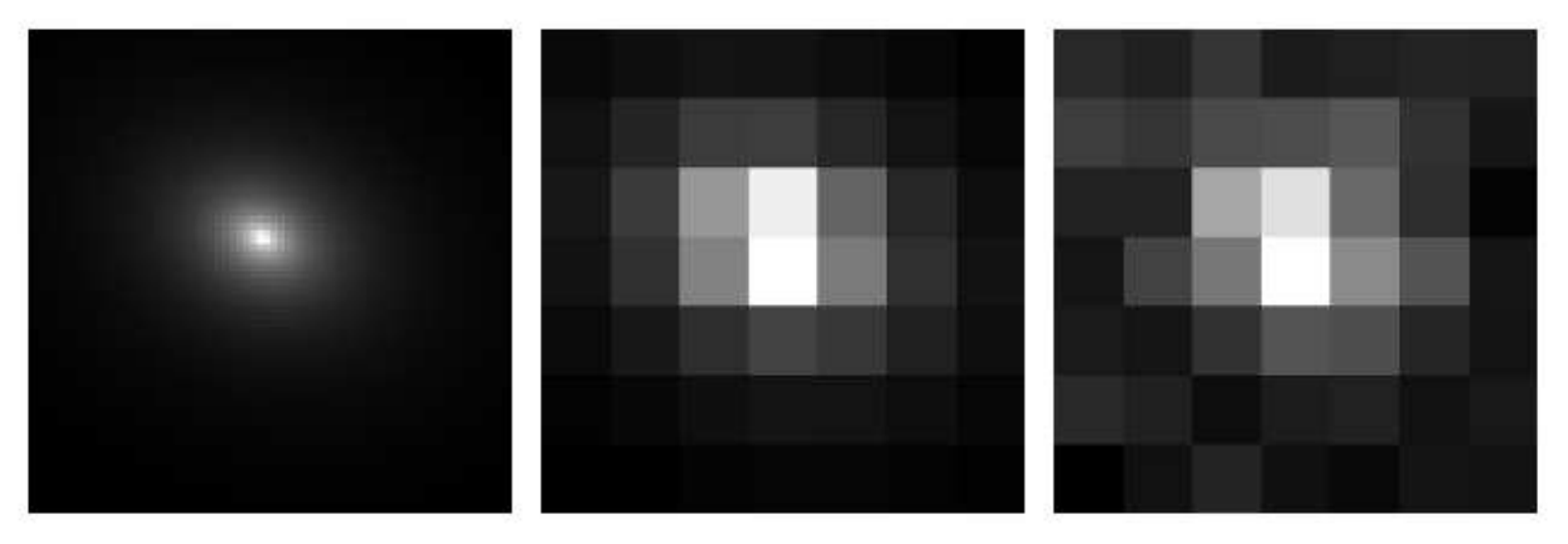}
\caption{An example of the simulated galaxy images used in this section. Here we have displayed the central region of the images, with the scale identical for each image. \emph{left panel}: First we simulated the galaxy using a high resolution grid consisting of $510\times510$ pixels. \emph{centre panel}: The pixelized noise free galaxy image is then produced by averaging over the pixel values in the high resolution grid to produce a grid of $51\times51$ pixels. \emph{right panel}: Gaussian noise is then added to the galaxy image with $\mathrm{SNR}=30$, in accordance with equation (\ref{eq:SNR}).}
\label{fig:gal_im}
\end{minipage}
\end{figure*} 
For the simulations in this section we follow the approach outlined in \cite{viola11} and consider sets of simulated galaxies assuming a S\'ersic brightness profile:
\begin{equation}\label{eq:sersic}
I(r)=I_0\exp\left[-b_{n_s}\left(\left(\frac{r}{R_e}\right)^{\frac{1}{n_s}}-1\right)\right],
\end{equation}
where $n_s$ is the S\'ersic index and where we assume that $n_s=1.5$, which is the average value for bright galaxies in the Cosmic Evolution Survey (COSMOS) field \citep{sargent07}. $R_e$ is the half light radius and $b_{n_s}$ is a constant which depends on $n_s$. A value of $n_s=1.5$ gives $b_{n_s}=2.674$. $I_0$ is the surface brightness of the galaxy at the half light radius. The size of each image is $51\times51$ pixels, with each galaxy model 10-fold oversampled, and we set $R_e=2$ pixels.

When background noise is introduced into the images, we assume a Gaussian noise, and fix the variance of the noise distribution, such that the resulting signal to noise ratio is 30. The signal to noise ratio (SNR) is defined as \citep{bridle10}
\begin{equation}\label{eq:SNR}
\mathrm{SNR}=\frac{\sqrt{\sum_{i=1}^NI_i^2}}{\sigma_b^2},
\end{equation}
where $I_i$ is the intensity in the $i^{\mathrm{th}}$ pixel of the low resolution image prior to the addition of noise and $\sigma_b$ is the dispersion in the background noise. An example of the process used to create the galaxy images in shown in Figure \ref{fig:gal_im}.

We simulate sets of galaxies using a Rayleigh intrinsic ellipticity distribution and then apply a constant shear to these galaxies. These sheared galaxies are then pixelized and Gaussian noise is added following the above procedure. We ignore the effects of PSF convolution and, when using a Gaussian weighting function to suppress noise at large scales, we set the width of the weighting function to $2R_e$. The effects of PSF convolution and the dependence of shear estimates on the size of the weighting function will be explored in future work. We recover estimates of the shear using a common variant of the KSB method and using the angle only method and compare the results. In order to use the angle only method we introduce a method of measuring the position angles of the galaxies using the light distribution of the galaxy images. 

We define the angular moments of the galaxy's surface brightness, $I(\bm{\theta})$, as
\begin{equation}\label{eq:moments}
Q_{ij...k}=\frac{\int\mathrm{d}^2\theta I(\bm{\theta})\theta_i\theta_i...\theta_k}{\int\mathrm{d}^2\theta I(\bm{\theta})},
\end{equation}
where $\bm{\theta}$ is the angular position of the image on the sky. We can define the ellipticity of the galaxy image in terms of the second order moments. The two definitions of ellipticity commonly used are
\begin{equation}\label{eq:eps_moments}
\epsilon=\frac{Q_{20}-Q_{02}+2iQ_{11}}{Q_{20}+Q_{02}+2\sqrt{Q_{20}Q_{02}-Q_{11}^2}},
\end{equation}
which corresponds to the shear transformation given in equation (\ref{eq:obs_ellip_shear}). An alternative form is
\begin{equation}\label{eq:chi_moments}
\chi=\frac{Q_{20}-Q_{02}+2iQ_{11}}{Q_{20}+Q_{02}},
\end{equation}
which is the form used in the KSB method and corresponds to a shear transformation, such that
\begin{equation}\label{eq:chi_shear}
\chi=\frac{\chi^{\mathrm{int}}+2g+g^2\chi^{\mathrm{int}^*}}{1+|g|^2+2\Re\left(g\chi^{\mathrm{int^*}}\right)}.
\end{equation}

When estimating the shear using the KSB method we follow the method outlined in \cite{viola11}, and first discussed by \cite{kaiser95}, where the estimator is found to be
\begin{equation}\label{eq:KSB_est}
\hat{g}_{\alpha}=\left<\left(P^{\mathrm{sh}}_{\alpha\beta}\right)^{-1}\chi^{\beta}\right>,
\end{equation}
such that the tensor, $P_{\alpha\beta}^{\mathrm{sh}}$, is approximated by half its trace

To find the centroid of the galaxy image in the presence of noise, we first apply the weighting function centred on the brightest pixel. We then recalculate the centroid using equation (\ref{eq:moments}) and re-apply the weighting function centred on this new estimation of the centroid. We iterate this step until the difference between successive estimates is less than $10^{-4}$ of a pixel.
\subsection{Using the angle only method}
\label{subsec:angles}
The $F_1\left(\left|\bm{g}\right|\right)$ function corresponding to the definition of ellipticity given in equation (\ref{eq:eps_moments}) is given in equation (\ref{eq:general_F}). It is also possible to derive the $F_1\left(\left|\bm{g}\right|\right)$ function which corresponds to the $\chi$-ellipticity definition given in equation (\ref{eq:chi_shear}), $F_1^{\chi}\left(\left|\bm{g}\right|\right)$. The form of this function is 
\begin{align}\label{eq:F_chi}
F_1^{\chi}\left(\left|\bm{g}\right|\right)=&\frac{1}{\pi}\int_0^{\left|\bm{\chi}_{\mathrm{max}}^{\mathrm{int}}\right|}\mathrm{d}\alpha^{\mathrm{int}}\mathrm{d}\left|\mathbf{\chi}^{\mathrm{int}}\right|f\left(\left|\bm{\chi}^{\mathrm{int}}\right|\right)\nonumber\\
&\times h_1^{\chi}\left(\left|\bm{g}\right|,\left|\bm{\chi}^{\mathrm{int}}\right|,\alpha^{\mathrm{int}}\right),
\end{align}
where $\alpha^{\mathrm{int}}$ is the intrinsic position angle and the function $h_1^{\chi}\left(\left|\bm{g}\right|,\left|\bm{\chi}^{\mathrm{int}}\right|,\alpha^{\mathrm{int}}\right)$ is
\begin{equation}\label{eq:chi_h1}
h_1^{\chi}\left(\left|\bm{g}\right|,\left|\bm{\chi}^{\mathrm{int}}\right|,\alpha^{\mathrm{int}}\right)=\frac{\chi_1'}{\sqrt{\chi_1'^2+\chi_2'^2}},
\end{equation}
with
\begin{align}\label{eq:chi_dash}
\chi_1'=&2\left|\bm{g}\right|+\left(1+\left|\bm{g}\right|^2\right)\left|\chi^{\mathrm{int}}\right|\cos\left(2\alpha^{\mathrm{int}}\right),\nonumber\\
\chi_2'=&\left(1-\left|\bm{g}\right|^2\right)\left|\chi^{\mathrm{int}}\right|\sin\left(2\alpha^{\mathrm{int}}\right).
\end{align}
We use this form of the $F_1\left(\left|\bm{g}\right|\right)$ function in the following analysis in order to make a direct comparison of the shear estimates recovered when using an angle only method - where we measure the position angles directly from the image data with those obtained using the KSB method. For this analysis we also assume a perfect knowledge of the intrinsic ellipticity distribution, $f\left(\left|\bm{\chi}^{\mathrm{int}}\right|\right)$; the effects of an imperfect knowledge of $f\left(\left|\bm{\chi}^{\mathrm{int}}\right|\right)$ are discussed in Section \ref{sec:errors_pofe}.

An estimation of the shear is obtained, such that 
\begin{align}\label{eq:shear_est_chi}
F_1^{\chi}\left(\left|\hat{\bm{g}}\right|\right)=&\sqrt{\left[\frac{1}{N}\sum_{i=1}^N\cos\left(2\alpha^{(i)}\right)\right]^2+\left[\frac{1}{N}\sum_{i=1}^N\sin\left(2\alpha^{(i)}\right)\right]^2},\nonumber\\
\hat{\alpha}_0=&\frac{1}{2}\tan^{-1}\left(\frac{\sum_{i=1}^N\sin\left(2\alpha^{(i)}\right)}{\sum_{i=1}^N\cos\left(2\alpha^{(i)}\right)}\right).
\end{align}

\subsection{A method for measuring the position angles}
\label{subsec:method_angles}
We now consider a method of obtaining an estimate of the position angle by considering the intensity profile of the galaxy as a function of the assumed position angle, under the assumption that the galaxy exhibits reflection symmetry about its major axis.

Given a noisy pixelized galaxy image, we begin by obtaining an estimate of the centroid. To find the centroid of the galaxy image we follow a similar procedure to that discussed above. First we multiply by the weighting function centred on the brightest pixel, however, for the angle only method we then convolve the image with a Gaussian kernel to reduce pixelization effects (the advantage of this step is discussed shortly), and then recalculate the centroid using equation (\ref{eq:moments}). We re-apply the weighting function to the original image centred on the new estimate of the centroid and convolve this image with the Gaussian kernel. We iterate this step until the difference between successive estimates is less than $10^{-4}$ of a pixel.

We then integrate the convolved, weighted surface brightness, $I_w(r,\alpha)$, over the radial direction, such that
\begin{equation}\label{eq:brightness_marg}
I'(\theta)=\int\mathrm{d}r I_w(r,\theta),
\end{equation}  
where $r=0$ corresponds to the centroid of the galaxy image and where $\theta$ is the assumed galaxy orientation. This gives us the integrated light distribution as a function of $\theta$. 

\begin{figure}
\centering
\includegraphics{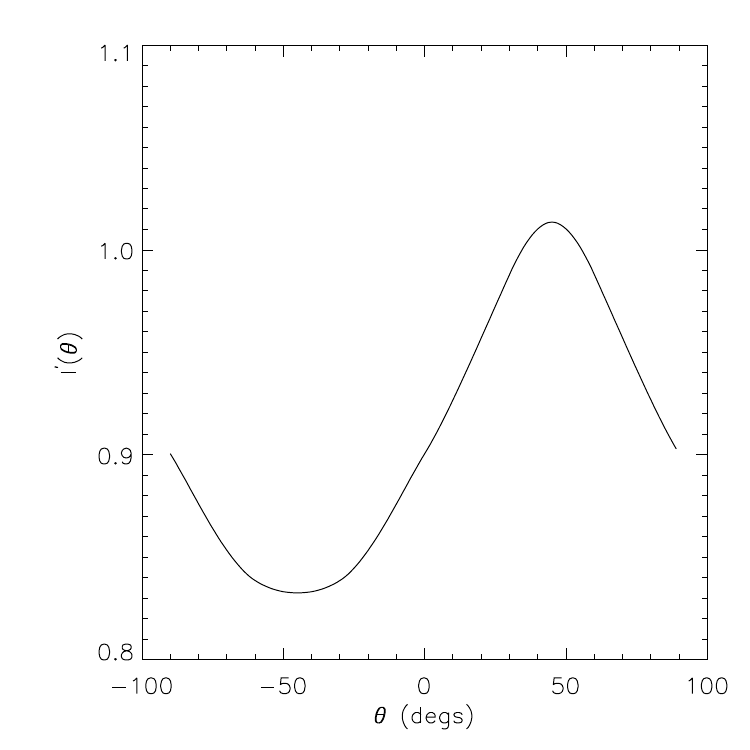}
\caption{The integrated light distribution as a function of the assumed galaxy orientation for the ideal case of zero noise and with the centroid of the galaxy situated at the centre of a pixel. The input position angle is $45^{\circ}$.}
\label{fig:I_alpha_dist}
\end{figure} 
Defining a set of axes which align with the edges of the pixels, we propose a method of performing this integration by rotating the image about the centroid and integrating along the x-axis of the image, through the centroid, such that
\begin{equation}\label{eq:brightness_x}
I'(\theta)=\int_{x_{min}}^{x_{max}}\mathrm{d}x I_{\theta}(x,\hat{y}),
\end{equation}
where $I_{\theta}$ is the convolved, weighted surface brightness distribution rotated by $-\theta$ about the centroid, and $\hat{y}$ is the $y$ component of the estimated centroid, $(\hat{x},\hat{y})$. This allows us to carry out the integration by simply summing the intensity over the row of pixels with $y=\hat{y}$.

If we omit the convolution with a Gaussian kernel, an example of this distribution is shown in Figure \ref{fig:I_alpha_dist}, where we obtain the integrated light distribution for a galaxy with ellipticity $\left|\boldsymbol{\chi}\right|=0.27$, position angle $\alpha=45^{\circ}$ and situated at the centre of the central pixel with zero noise. Using this distribution we can obtain an estimate of the position angle of the galaxy such that
\begin{equation}\label{eq:alpha_est_dist}
\hat{\alpha}=\frac{1}{2}\tan^{-1}\left(\frac{\int\mathrm{d}\theta I'\left(\theta\right)\sin\left(2\theta\right)}{\int\mathrm{d}\theta I'\left(\theta\right)\cos\left(2\theta\right)}\right).
\end{equation}
\begin{figure*}
\begin{minipage}{6in}
\centering
\includegraphics{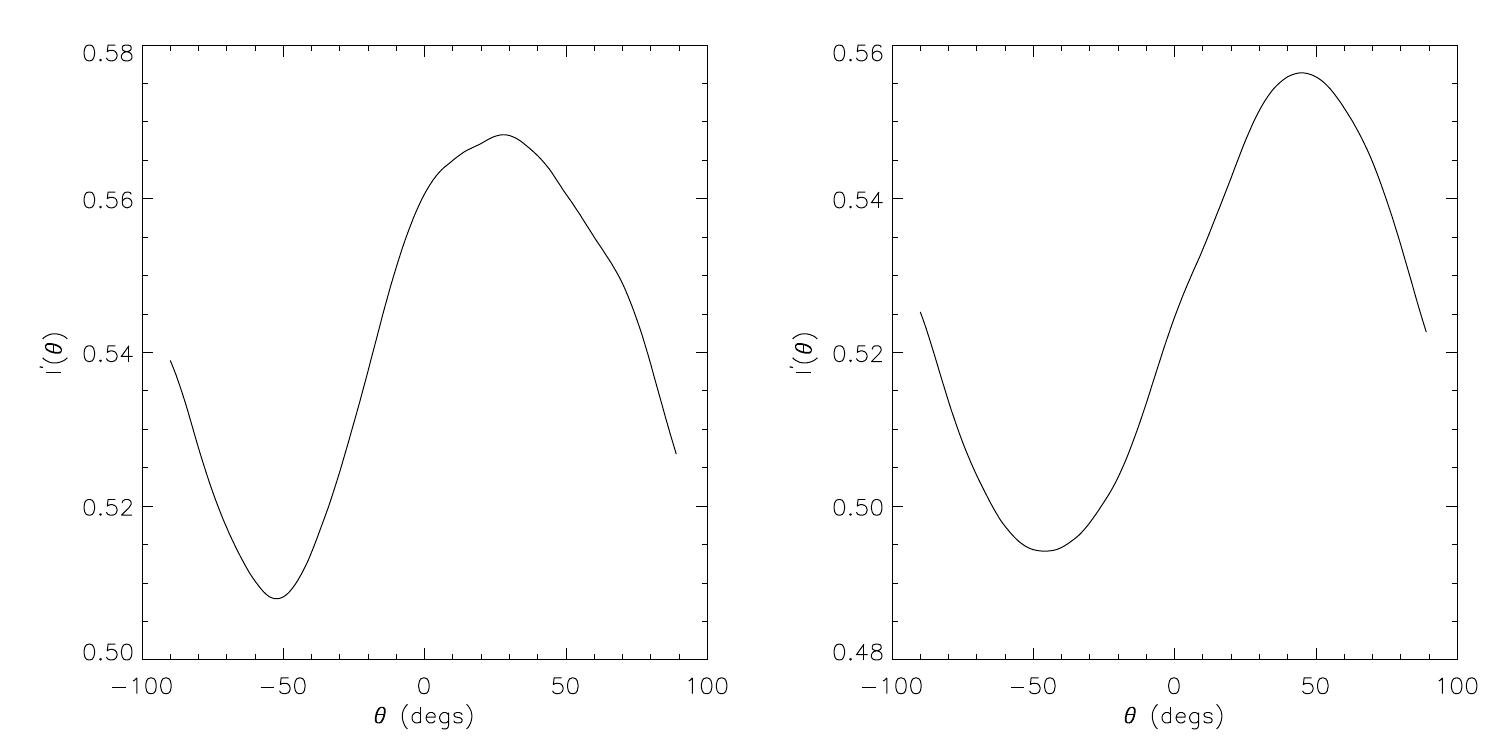}
\caption{\emph{left panel}: The integrated light distribution as a function of the assumed galaxy orientation for the case of a randomly positioned galaxy. This distribution leads to an incorrect estimation of the true position angle. \emph{right panel}: The integrated light distribution once the image has been convolved with a Gaussian kernel of width 1 pixel; this reduces the effects of pixelization }
\label{fig:I_alpha_dist_rc_nc_c}
\end{minipage}
\end{figure*} 

Applying this equation to the above distribution leads to the exact estimate of $\hat{\alpha}=45^{\circ}$. However, if we now consider the case where the centroid of the galaxy is not positioned at the centre of a pixel, the integrated light distribution is not so well behaved. The left-hand panel in Figure \ref{fig:I_alpha_dist_rc_nc_c} shows one particular distribution using a randomly centred galaxy. This distribution leads to an estimated position angle of $31.0^{\circ}$ for an input angle of $45^{\circ}$. This error in the estimation is due to the pixelization of the image. One can soften the effect of pixelization by convolving the image with a Gaussian kernel as described above. After we have done this with a kernel of width one pixel, we obtain the distribution shown in the right-hand panel of Figure \ref{fig:I_alpha_dist_rc_nc_c}. Using this distribution we obtain the estimate of $\hat{\alpha}=45.1^{\circ}$. For the remainder of this section we use a Gaussian kernel of width one pixel in every position angle measurement performed.

\subsection{Debiasing angle only shear estimates using simulations}
\label{subsec:debias_angles}
For the analysis in Section \ref{sec:method} we assumed that measurement errors on the position angles were independent of the true position angles. However, for a fixed signal to noise ratio we find that there are two sources of non-zero covariance between the measurement errors and the position angles, these are pixelization and a correlation between the measurement errors and galaxy ellipticities in the presence of a non-zero shear. To understand why these covariances present a problem we begin by writing the mean estimated cosines and sines of the measured position angles, $\hat{\alpha}$, in terms of the true position angles, $\alpha$, and an error on the measurement, $\delta\alpha$, such that
\begin{align}\label{eq:trig_estimates_full}
\left<\cos\left(2\hat{\alpha}\right)\right>=&\left<\cos\left(2\alpha+2\delta\alpha\right)\right>\nonumber\\
\left<\sin\left(2\hat{\alpha}\right)\right>=&\left<\sin\left(2\alpha+2\delta\alpha\right)\right>.
\end{align}
If we now abandon the assumption that the measurement errors are independent of the position angles of the galaxies and make no assumption about the distribution of the errors, we can write equation (\ref{eq:trig_estimates_full}) in terms of the covariance between the true position angles and the measurement errors, such that
\begin{align}\label{eq:cov_c_s_relation}
C'=&\left<\cos\left(2\alpha\right)\right>\beta_c-\left<\sin\left(2\alpha\right)\right>\beta_s\nonumber\\
S'=&\left<\sin\left(2\alpha\right)\right>\beta_c+\left<\cos\left(2\alpha\right)\right>\beta_s,
\end{align}
where we define
\begin{align}\label{eq:cov_c_s}
C'=&\left<\cos\left(2\hat{\alpha}\right)\right>-\mathrm{cov}\left(\cos\left(2\alpha\right),\cos\left(2\delta\alpha\right)\right)+\nonumber\\
&\mathrm{cov}\left(\cos\left(2\alpha\right),\cos\left(2\delta\alpha\right)\right),\nonumber\\
S'=&\left<\sin\left(2\hat{\alpha}\right)\right>-\mathrm{cov}\left(\sin\left(2\alpha\right),\cos\left(2\delta\alpha\right)\right)+\nonumber\\
&\mathrm{cov}\left(\cos\left(2\alpha\right),\sin\left(2\delta\alpha\right)\right),\nonumber\\
\beta_c=&\left<\cos\left(2\delta\alpha\right)\right>,\nonumber\\
\beta_s=&\left<\sin\left(2\delta\alpha\right)\right>.
\end{align}
We can now use equation (\ref{eq:cov_c_s_relation}) to write the means of the cosines and sines of the true position angles as
\begin{align}\label{eq:mean_c_s}
\left<\cos\left(2\alpha\right)\right>=&\frac{1}{\beta_c^2+\beta_s^2}\left(C'\beta_c+S'\beta_s\right),\nonumber\\
\left<\sin\left(2\alpha\right)\right>=&\frac{1}{\beta_c^2+\beta_s^2}\left(S'\beta_c-C'\beta_s\right).
\end{align}
It is these averages which must be used to estimate the shear when using equation (\ref{eq:shear_est_chi})\footnote{This result is general and is expected to hold for any method used to measure galaxy position angles.}. Incorrectly accounting for the covariance terms in equation (\ref{eq:mean_c_s}) will introduce a bias into the shear estimates.

To understand how pixelization of the galaxy images introduces non-zero covariance terms into equation (\ref{eq:mean_c_s}), let us assume that the centre of a galaxy coincides with the centre of a pixel. In the absence of noise the central pixel of this image will be the brightest. If we define the pixel axes as a set of axes with the origin at the centre of the brightest pixel and with the axes aligned with the edges of the pixel, then the contribution of light from the central pixel will be greatest in the directions of $\pm45^{\circ}$ and least in the directions of $0^{\circ}$ and $90^{\circ}$. This effect will bias the angle measurements towards angles of $\pm45^{\circ}$ and away from angles of $0^{\circ}$ and $90^{\circ}$. Convolving the galaxy image with a Gaussian kernel reduces this effect. However, for a randomly positioned galaxy centroid, with a signal to noise ratio of 30, using a Gaussian kernel with a width of 1 pixel results in correlations on the order of $1\%$, which produces significant biases in the shear estimates. The covariance terms in equation (\ref{eq:cov_c_s}) due to this effect can be obtained using a set of simulated galaxy images under the assumption of zero shear. From these simulations we can also obtain a first estimate of $\beta_c$ and $\beta_s$. If we align the shear axes with the pixel axes, then the $g_1$ component of the shear will be aligned with the pixel axes and the $g_2$ component will align with the directions of $\pm45^{\circ}$, this will lead to a biasing in the direction of $g_2$ and away from the direction of $g_1$. In order to avoid this effect, we choose to orientate our shear axes, when performing the angle only analysis, such that the direction of the $g_1$ component is orientated at $-22.5^{\circ}$ to the pixel axes.

The second source of bias arises from correlations between measurement errors and galaxy ellipticities in the presence of a non-zero shear. For a fixed signal to noise ratio it is found that the measurement error on the position angle is dependent on the modulus of the ellipticity - galaxies with high ellipticities have smaller measurement errors on the position angles than galaxies with low ellipticities. In the absence of a shear there will be no correlation between orientation and ellipticity and, hence, there will be no bias contribution due to this effect. However, in the presence of a non-zero shear there will be a preference for galaxies with a higher ellipticity to align with the direction of the shear. This implies that, on average, the measurement errors on galaxy position angles where the galaxies are aligned with the shear are smaller than those where the galaxies are anti-aligned with the shear. This effect is also large enough to produce significant biases in the shear estimates. In the following subsection we introduce an iterative method using simulations to significantly reduce this effect. 
\begin{figure*}
\begin{minipage}{6in}
\centering
\includegraphics{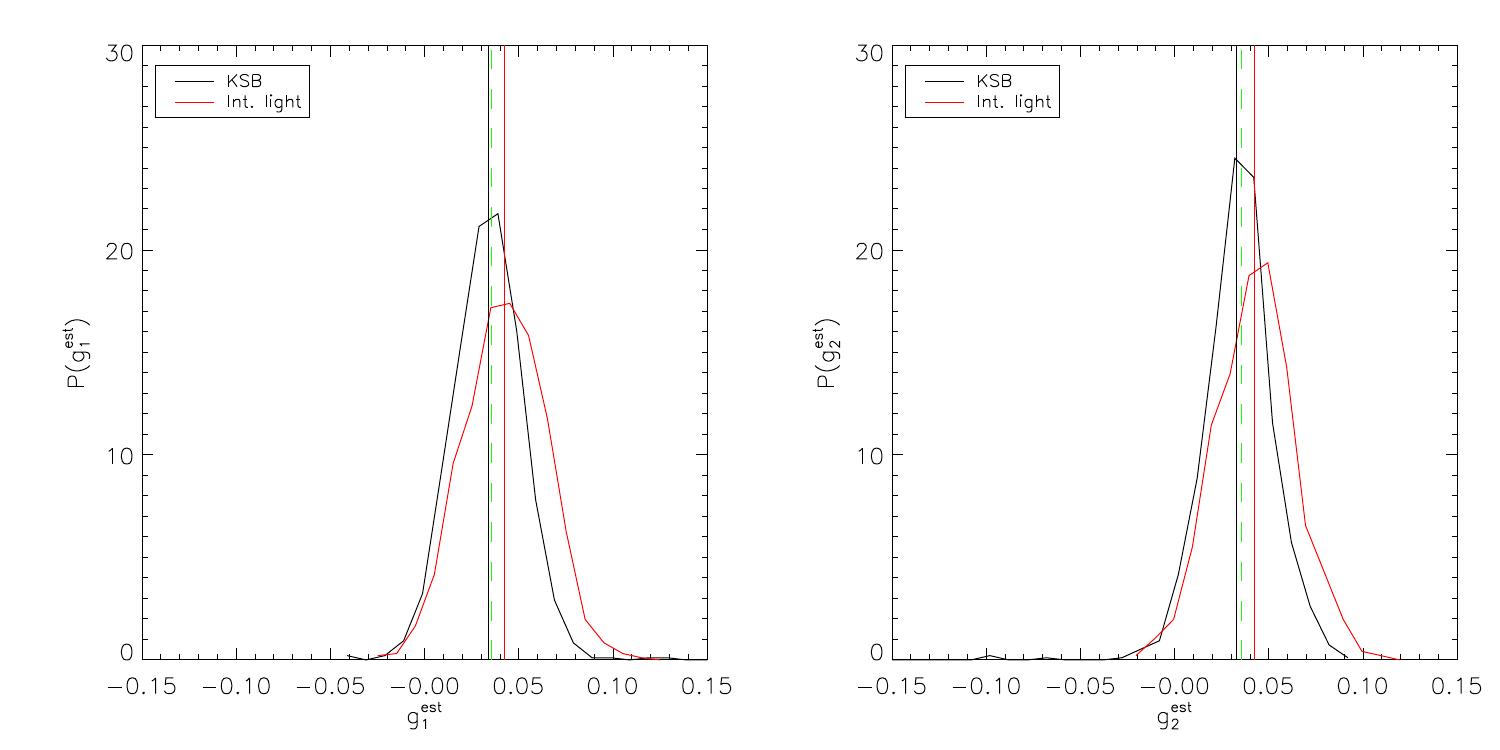}
\caption{The distribution of the estimates for the two components of the shear including correction for pixelization. The black curves show the distributions in shear estimates using the KSB method, with the vertical black line indicating the mean estimate. The red curves show the distributions in shear estimates using the angle only method, where we measure the position angles of the galaxies using the integrated light distribution and use sets of zero-shear realizations in order to reduce the bias due to pixelization; the vertical red line indicates the mean estimate. The green dashed line indicates the input shear signal. The residual bias in the angle only estimates is due to correlations between the measurement errors and the position angles which result from a non-zero shear signal.}
\label{fig:hist_noit}
\end{minipage}
\end{figure*} 
\subsection{A comparison of angle only shear estimates with the KSB method}
\label{subsec:compare_KSB_angles}
Using sets of simulated galaxy images, with a signal to noise ratio of 30, assuming a Rayleigh intrinsic ellipticity distribution with a dispersion $\sigma_{\chi}=0.3/\sqrt{2}$, using an input shear signal of $g_1=g_2=0.05/\sqrt{2}$ and with 100 galaxies per realization, we recovered shear estimates from 960 realizations. We compared the estimates recovered using the angle only estimator, where the angles were measured using the method described above, with those recovered using the KSB method. For the angle only method we also used 960 zero-shear realizations in order to obtain a first estimate of the covariance terms in equation (\ref{eq:mean_c_s}); this corrects for the effect of pixelization on the shear estimates. The results of these simulations are shown in Figure \ref{fig:hist_noit}. The two panels show the distribution of estimates for the two components of the shear. The black curves show the distributions of recovered shear estimates using the KSB method. The red curves show the distributions of recovered shear estimates using the angle only method. Here we see that the width of the distribution is slightly larger for the angle only method. Also, the bias in the shear estimates using the angle only method is clearly visible. As explained above, this bias is due to the correlation between measurement errors and galaxy ellipticities, which is introduced in the presence of a non-zero shear. However, one can obtain a better estimate of these covariance terms by using the estimated shear values as the input shear for a new set of simulated galaxy images. 
\begin{figure*}
\begin{minipage}{6in}
\centering
\includegraphics{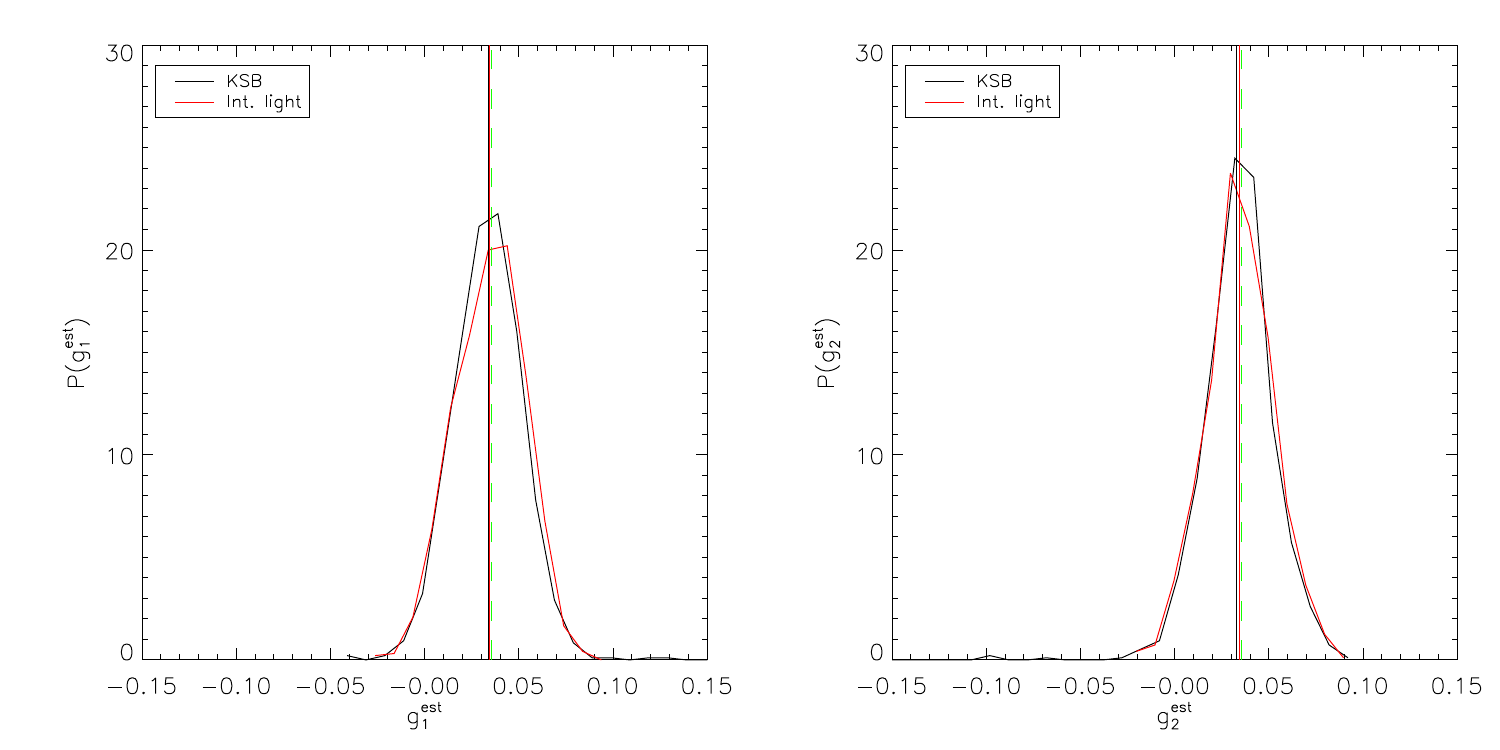}
\caption{The distribution in shear estimates for the two components of the shear including corrections for pixelization and an iterative scheme for removing the bias due to shape dependent noise. The black curves show the distributions in shear estimates using the KSB method. The red curves show the distributions in shear estimates using the angle only method, where we have used the estimated shear values shown in Figure \ref{fig:hist_noit} as the input shear signal in a set of simulations, which we then use to re-evaluate the covariance terms in equation (\ref{eq:cov_c_s}). The bias in the angle only shear estimates has been greatly reduced.}
\label{fig:hist_it}
\end{minipage}
\end{figure*} 

Using each of the shear estimates shown in Figure \ref{fig:hist_noit} as the input shear for a further 960 realizations, we re-evaluated the covariance and $\beta$ terms in equation (\ref{eq:cov_c_s}). We then used these values of the covariances to obtain new estimates of the shear. The distribution in these estimates is shown in Figure \ref{fig:hist_it} as the red curve. From these plots we see that the bias in the shear estimates has been greatly reduced. We also see that the width of the distribution of the shear estimates using the angle only method has been reduced, and is now similar to the width of the distribution of shear estimates using the KSB method.  
\begin{figure*}
\begin{minipage}{6in}
\centering
\includegraphics{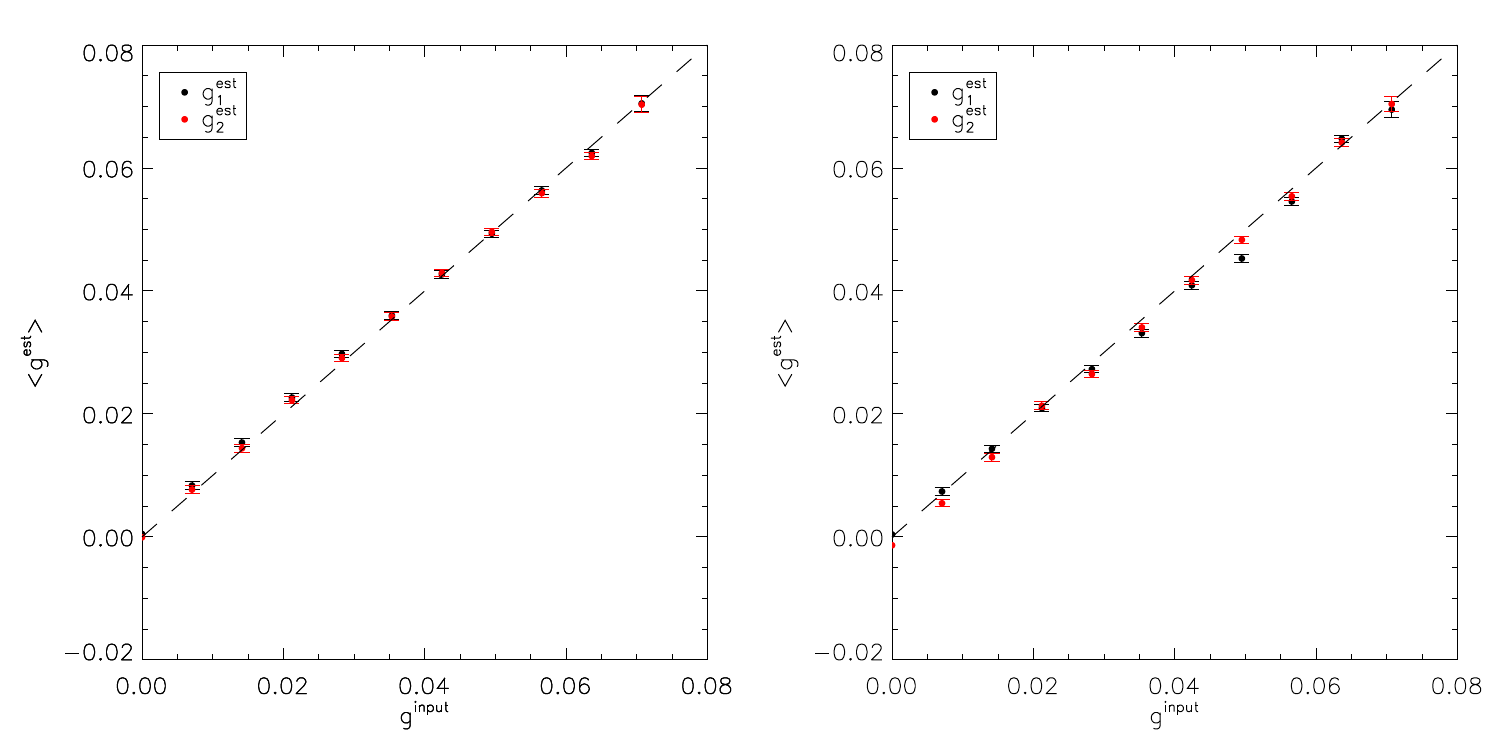}
\caption{The recovered shear estimates as a function of the input shear values. The \emph{left panel} shows the recovered shear values obtained using the angle only method, and upon using one iteration in order to calculate the covariance terms in equation (\ref{eq:cov_c_s}). The \emph{right panel} shows the recovered shear values using the KSB method.}
\label{fig:gvg}
\end{minipage}
\end{figure*} 

Next, we carried out this analysis for a range of input $\left|\bm{g}\right|$ values, keeping the input shear position angle fixed at $\alpha_0=22.5^{\circ}$. The results of this test are shown in Figure \ref{fig:gvg}. The left panel shows the results of the angle only method upon carrying out the iterative procedure described above, using just one iteration. The right panel shows the results of the KSB method. Each point plotted consists of 960 realizations, with each realization consisting of 100 galaxies. From these plots, we see that the performance of the angle only method is comparable with the KSB method, and that the distribution of the recovered shear values is consistent with noise for the number of galaxies and realizations used. 

\section{Impact of errors on $\MakeLowercase{f}\left(\left|\bm{\epsilon}^{\mathrm{\MakeLowercase{int}}}\right|\right)$ estimates}
\label{sec:errors_pofe}
All of the work prior to this section has assumed an exact knowledge of $f\left(\left|\bm{\epsilon}^{\mathrm{int}}\right|\right)$. In practice an estimate of $f\left(\left|\bm{\epsilon}^{\mathrm{int}}\right|\right)$ is necessary in order to allow an estimation of the shear using the angle only method. Errors on the form of $f\left(\left|\bm{\epsilon}^{\mathrm{int}}\right|\right)$ will, therefore, propagate as a bias into estimates of $\left|\bm{g}\right|$. In this section we examine the effects of the size of the sample of galaxies, and of measurement errors on the ellipticities of galaxies in the sample, on estimates of the shear. For the following analysis we assumed the underlying form of $f\left(\left|\bm{\epsilon}^{\mathrm{int}}\right|\right)$ to be that given in equation (\ref{eq:CFHT_dist}).
\begin{figure*}
\begin{minipage}{6in}
\centering
\includegraphics{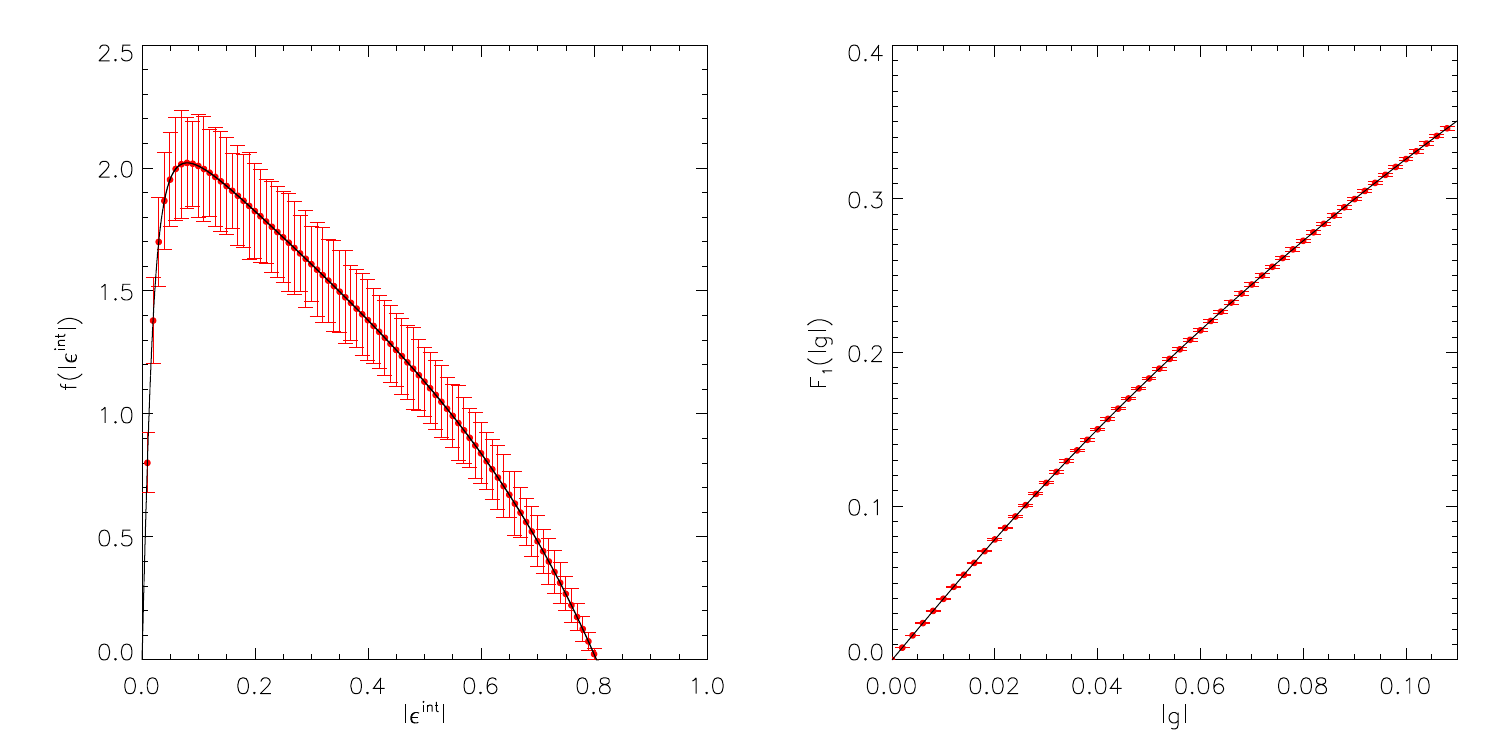}
\caption{\emph{left panel}: The $f\left(\left|\bm{\epsilon}^{\mathrm{int}}\right|\right)$ for the case of $5\times10^{4}$ galaxies. The red points indicate the mean of $f\left(\left|\bm{\epsilon}^{\mathrm{int}}\right|_i\right)$ recovered from 100 realizations. The error bars indicate the standard deviation of the values of $f\left(\left|\bm{\epsilon}^{\mathrm{int}}\right|_i\right)$ over 100 realizations (note that only 1 in 10 points have been plotted for clarity). The black curve shows the true $f\left(\left|\bm{\epsilon}^{\mathrm{int}}\right|\right)$. \emph{right panel} The estimated $F_1\left(\left|\bm{g}\right|\right)$ function obtained from the estimated $f\left(\left|\bm{\epsilon}^{\mathrm{int}}\right|\right)$. The red points show the mean estimated $F_1\left(\left|\bm{g}\right|_i\right)$ over 100 realizations. The error bars indicate the standard deviation over 100 realizations (note that only 1 in 20 points have been plotted for clarity). The black curve shows the true $F_1\left(\left|\bm{g}\right|\right)$ function obtained using the true $f\left(\left|\bm{\epsilon}^{\mathrm{int}}\right|\right)$.}
\label{fig:pe_FT_50000}
\end{minipage}
\end{figure*} 

We began by assuming a negligible measurement error on the galaxy ellipticities. We are thus implicitly assuming that a high signal to noise sample of galaxies are available from which $f\left(\left|\bm{\epsilon}^{\mathrm{int}}\right|\right)$ can be estimated with negligible measurement errors. In order to explore how the size of the galaxy sample can be used to estimate how $f\left(\left|\bm{\epsilon}^{\mathrm{int}}\right|\right)$ affects the shear estimates, we reconstructed $f\left(\left|\bm{\epsilon}^{\mathrm{int}}\right|\right)$ using sample sizes consisting of a various number of galaxies, $N$ by producing histograms of $\left|\epsilon^{\mathrm{int}}\right|$ which were randomly drawn from the underlying distribution. For each sample size we repeated this process for 100 realizations.

For each estimate of $f\left(\left|\bm{\epsilon}^{\mathrm{int}}\right|\right)$ we calculated the corresponding $F_1\left(\left|\bm{g}\right|\right)$ function, $\hat{F}_1\left(\left|\bm{g}\right|\right)$, the results of this analysis are shown, for the case of $5\times10^{4}$ galaxies, in Figure \ref{fig:pe_FT_50000}. From this plot we see that the errors on the estimates of $F_1\left(\left|\bm{g}\right|\right)$ are small compared to the errors on the $f\left(\left|\bm{\epsilon}^{\mathrm{int}}\right|\right)$, this is because the calculation of the $F_1\left(\left|\bm{g}\right|\right)$ function involves an integral over the estimated values of $f\left(\left|\bm{\epsilon}^{\mathrm{int}}\right|_i\right)$, which smooths the $f\left(\left|\bm{\epsilon}^{\mathrm{int}}\right|\right)$.
\begin{figure}
\centering
\includegraphics{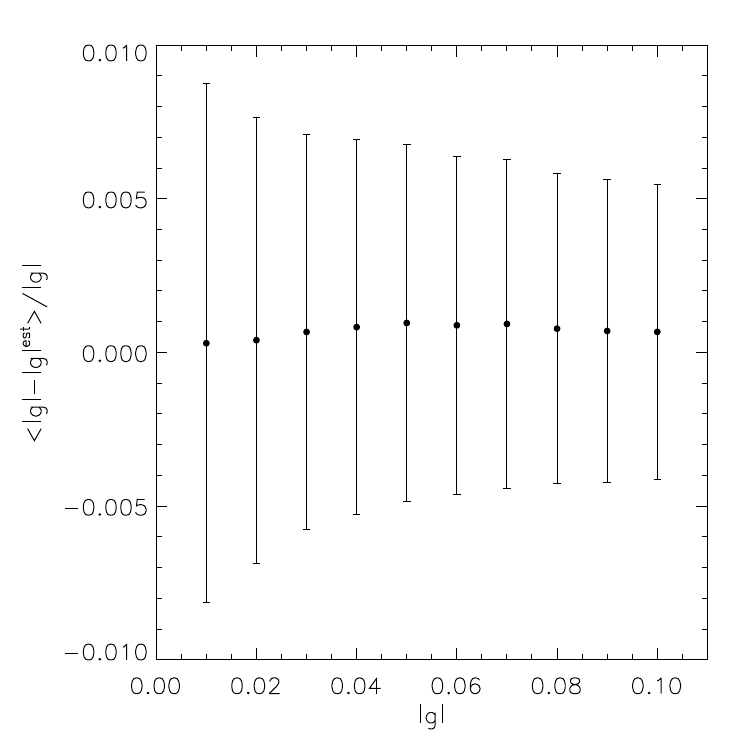}
\caption{The fractional bias in the recovered $\left|\bm{g}\right|$ estimates for the case of $5\times10^{4}$ galaxies. The error bars indicate the fractional standard deviation of the estimates over 100 realizations.}
\label{fig:mod_g_50000}
\end{figure} 

We propagated the errors on $\hat{F}_1\left(\left|\bm{g}\right|\right)$ into errors on the estimates of $\left|\bm{g}\right|$. This procedure was carried out for a variety of input shear signals, with the results displayed in Figure \ref{fig:mod_g_50000}.
\begin{figure}
\centering
\includegraphics{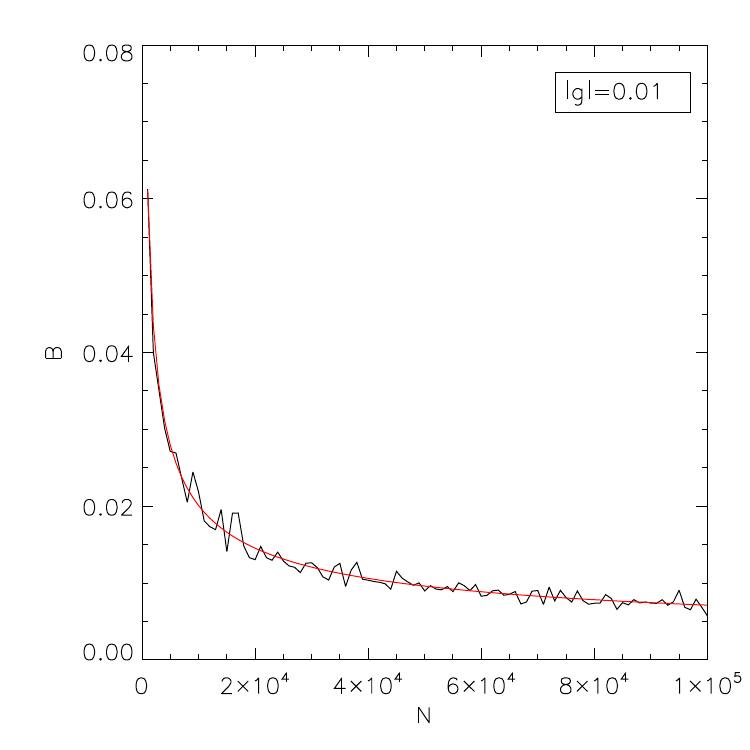}
\caption{The absolute value of the fractional bias plus the fractional error (equation (\ref{eq:frac_bias_pe})) as a function of the number of galaxies in the sample used to estimate $f\left(\left|\bm{\epsilon}^{\mathrm{int}}\right|\right)$. The black curve shows the values obtained from the simulations. Over-plotted in red is a curve of the form $M/\sqrt{N}+C$.}
\label{fig:bias_plot_N_noerr}
\end{figure} 

In order to quantify the errors introduced due to an imprecise knowledge of $f\left(\left|\bm{\epsilon}^{\mathrm{int}}\right|\right)$ we can, for example, constrain the number of galaxies in the sample used to estimate $f\left(\left|\bm{\epsilon}^{\mathrm{int}}\right|\right)$ such that biases in the estimations of $\left|\bm{g}\right|$ are  below some threshold value at a particular confidence level. As an example, let us define
\begin{equation}\label{eq:frac_bias_pe}
B\equiv\frac{\left|\left<\left|\bm{g}\right|-\left|\hat{\bm{g}}\right|\right>\right|+\sigma_{\left|\hat{\bm{g}}\right|}}{\left|\bm{g}\right|},
\end{equation}
which is the fractional absolute value of the bias plus the fractional standard deviation of the recovered estimates of $\left|\bm{g}\right|$. We can find the number of galaxies required to estimate $f\left(\left|\bm{\epsilon}^{\mathrm{int}}\right|\right)$, such that $B$ is below some value for a specific set of $\left|\bm{g}\right|$ values. From Figure \ref{fig:mod_g_50000} we see that the range of the fractional error bars on the recovered estimates of $\left|\bm{g}\right|$ decreases as the input $\left|\bm{g}\right|$ increases, therefore, for the analysis which follows we focused our attention on the case when $\left|\bm{g}\right|=0.01$.

In Figure \ref{fig:bias_plot_N_noerr} we have plotted $B$ as a function of the number of galaxies in the sample (black curve). We see, as expected, that the value of $B$ decreases as the number of galaxies increases. For a fixed bin size this is a result of the fractional error decreasing as the number of galaxies increases. The red curve in Figure \ref{fig:bias_plot_N_noerr} is a curve of the form $M/\sqrt{N}+C$, where $M$ and $C$ are parameters fitted to the data. For this data we find $M\approx1.90$ and $C\approx1.13\times10^{-3}$. The additional constant, $C$, quantifies the residual bias due to the finite bin width used when reconstructing $f\left(\left|\bm{g}\right|\right)$.

From the fitted curve we can calculate the number of galaxies that one would require in order to ensure that $B$ is less than some value. We have verified that the recovered estimates of $\left|\bm{g}\right|$ are approximately Gaussian distributed about the mean estimate. We can, therefore, state that the bias in the estimates of $\left|\bm{g}\right|$ arising as a result of a finite number of galaxies in the sample will be less than or equal to $1.0\%$ for all values of $\left|\bm{g}\right|\ge0.01$, at a confidence level of $68\%$, if $B=0.01$. We can then invert the fitted curve in Figure \ref{fig:bias_plot_N_noerr} to find that, for this constraint, we need $N\gsim5\times10^{4}$. Looking again at Figure \ref{fig:mod_g_50000}, which displays the fractional bias for the case when $N=5\times10^{4}$ we see that the error bars are, indeed, contained within the range $[-0.01, 0.01]$.

For any true ellipticity measurements there will, of course, be measurement errors. These errors modify the form of the estimated $f\left(\left|\bm{\epsilon}^{\mathrm{int}}\right|\right)$ by distributing the measured $\left|\bm{\epsilon}^{\mathrm{int}}\right|$ more evenly between the bins. This is true even if the measurements of $\left|\bm{\epsilon}^{\mathrm{int}}\right|$ are unbiased.
\begin{figure}
\centering
\includegraphics{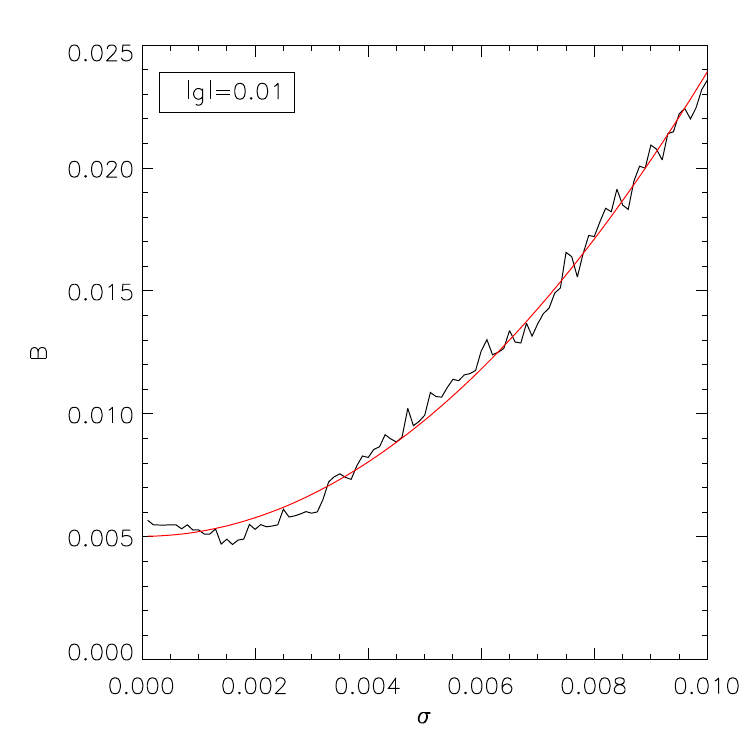}
\caption{The absolute value of the fractional bias plus the fractional error, for the case of $2\times10^5$ galaxies, as a function of the ellipticity measurement error. The black curve shows the values obtained from the simulations. Over-plotted in red is a curve of the form $M\sigma^2+C$.}
\label{fig:bias_plot_sig_err}
\end{figure} 

Assuming $2\times10^{5}$ galaxies in the sample used to estimate $f\left(\left|\bm{\epsilon}^{\mathrm{int}}\right|\right)$, we repeated the analysis above. This time we added a Gaussian measurement error to the components of the ellipticity before estimating $\left|\bm{\epsilon}^{\mathrm{int}}\right|$. To achieve this, we first randomly draw $2\times10^{5}$ samples of $\left|\bm{\epsilon}^{\mathrm{int}}\right|_i$ from the underlying ellipticity distribution. We then simulated the measured components of the ellipticity, such that
\begin{align}
\hat{\epsilon}^{\mathrm{int}}_1=&\left|\bm{\epsilon}^{\mathrm{int}}\right|\cos\left(2\alpha^{\mathrm{int}}\right)+\delta_{\epsilon_1},\nonumber\\
\hat{\epsilon}^{\mathrm{int}}_2=&\left|\bm{\epsilon}^{\mathrm{int}}\right|\sin\left(2\alpha^{\mathrm{int}}\right)+\delta_{\epsilon_2},
\end{align}
where $\alpha^{\mathrm{int}}$ is uniform distributed in the range $-\pi/2<\alpha^{\mathrm{int}}\le\pi/2$ and where $\delta_{\epsilon_1}$ and $\delta_{\epsilon_2}$ are Gaussian distributed measurement errors with zero mean and variance $\sigma^2$. We used these noisy ellipticity values to calculate the measured $\left|\epsilon^{\mathrm{int}}\right|$ of the galaxies, such that $\left|\hat{\epsilon}^{\mathrm{int}}\right|=\sqrt{\hat{\epsilon}^{\mathrm{int}^2}_1+\hat{\epsilon}^{\mathrm{int}^2}_2}$. The $\left|\hat{\epsilon}^{\mathrm{int}}\right|_i$ were then binned to give the estimated $f\left(\left|\bm{\epsilon}^{\mathrm{int}}\right|\right)$. 

The number of galaxies in the sample was chosen, such that the error bars shown in figure \ref{fig:mod_g_50000} were reduced, approximately, by a factor of two. This reduction in the size of the error bars resulting from a finite sample size allows a value of $B\leq0.01$ to be achieved when measurement errors on the ellipticities are included.

Figure \ref{fig:bias_plot_sig_err} shows $B$ as a function of $\sigma$ for the case of $2\times10^{5}$ galaxies and for $\left|\bm{g}\right|=0.01$. Over plotted in red is a curve of the form $M\sigma^2+C$, and fitted with $M\approx189$ and $C\approx5.03\times10^{-3}$. We can now constrain the allowed value of $\sigma$ such that $B$ is less than some value. From the curve fitted in Figure \ref{fig:bias_plot_sig_err} we find that, for $2\times10^{5}$ galaxies, $B\le0.01$ for $\left|\bm{g}\right|\ge0.01$ if $\sigma\lesssim0.005$. Therefore, we conclude that galaxies with a high signal to noise must be used if we are to avoid significant biases in the shear estimates arising from an imperfect estimate of $f\left(\left|\bm{\epsilon}^{\mathrm{int}}\right|\right)$. 

\section{Tests on simulations}

\label{sec:sims}
\begin{figure*}
\begin{minipage}{6in}
\centering
\includegraphics{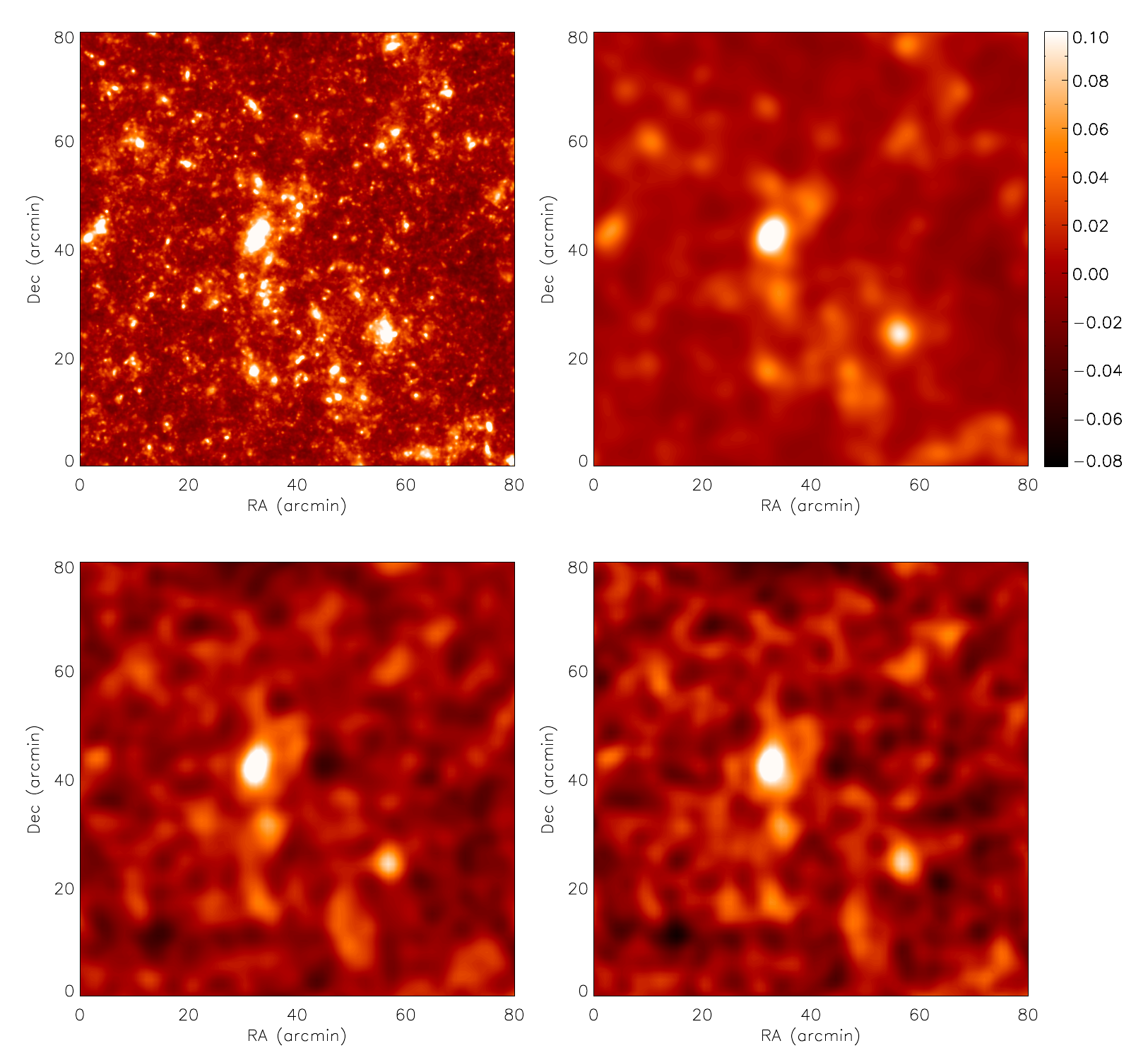}
\caption{Reconstruction of the distribution of dark matter in a
  $1.75 \text{ deg}^2$ region of the simulations. The upper-left panel
  shows the input convergence field, with the upper-right panel
  showing the input convergence field smoothed on a scale of 1.5
  arcmin. The lower left panel shows the simulated reconstruction
  using the standard method with full ellipticity information and with
  zero measurement errors. The lower right panel shows the
  reconstruction using the 3rd order estimator of the position angle
  only approach in the presence of measurement errors on the galaxy
  position angles with $\sigma_{\alpha}=15^{\circ}$.}
\label{fig:kappa_sim}
\end{minipage}
\end{figure*}

To compare the performance of the position angle only method with the
standard method based on full ellipticity measurements, we have
compared the convergence fields reconstructed from numerical
simulations using both approaches. These tests were performed
following the procedure outlined in \cite{brown11b}. Briefly, we used
a single field from the simulated lensing convergence and shear maps
of \cite{white05}, which consist of $\approx1000\text{ deg}^2$ of
simulated sky, based on a $\Lambda$CDM cosmology with the parameters:
$\Omega_m=0.28$, $\Omega_bh^2=0.024$, $h=0.7$, $\sigma_8=0.9$ and
$n_s=1$.
The input convergence distribution used for the simulation is shown in
the upper-left panel of Fig.~\ref{fig:kappa_sim}. The upper-right panel
shows the input convergence distribution smoothed on a scale of 1.5
arcmin.

We simulated a population of source galaxies assuming that the
intrinsic ellipticities, $\left|\bm{\epsilon}^{\mathrm{int}}\right|$,
follow the log-normal distribution
\begin{equation}\label{eq:log-normal}
f\left(\left|\bm{\epsilon}^{\mathrm{int}}\right|\right)=\frac{K}{\left|\bm{\epsilon}^{\mathrm{int}}\right|}\exp\left(-\frac{\left(\ln\left(\left|\bm{\epsilon}^{\mathrm{int}}\right|\right)-\mu\right)^2}{2\sigma^2}\right),
\end{equation}
where the mean and variance are given by
\begin{align}\label{eq:log-n-mean-var}
\text{mean}=&\exp\left(\mu+\frac{\sigma^2}{2}\right)\nonumber,\\
\sigma_{\epsilon}^2\equiv\text{variance}=&\left(\exp\left(\sigma^2\right)-1\right)\exp\left(2\mu+\sigma^2\right),
\end{align}
and where the normalization constant, $K$, was determined numerically.

In anticipation of our tests on the CFHTLenS data (see
Section~\ref{sec:CFHT}) we set the mean value of
$\left|\bm{\epsilon}^{\mathrm{int}}\right|$ to be 0.3370, which we
estimated from the CFHTLenS data by finding the average of the modulus
of the observed ellipticity value,
$\left<\left|\bm{\epsilon}^{\mathrm{obs}}\right|\right>$.  The
variance is taken to be $\sigma_{\epsilon}^2=0.2539^2$, which is the square of the
dispersion fitted to the disc dominated galaxies in the CFHTLenS
data. We also applied a maximum cut-off to the ellipticity of
$\left|\bm{\epsilon}^{\mathrm{int}}_{\mathrm{max}}\right|=0.804$
\citep{miller13}. It should be noted at this point that the
ellipticities of the CFHTLenS data are not well described by a
log-normal distribution. However, this distribution is easily
simulated and is useful in demonstrating our method.

The number density of the
background galaxies was taken to be $\bar{n}=17\text{ arcmin}^{-2}$
in accordance with the number density of resolved galaxy images
observed by the CFHTLenS \citep{heymans12}.
\begin{figure}
\centering
\includegraphics{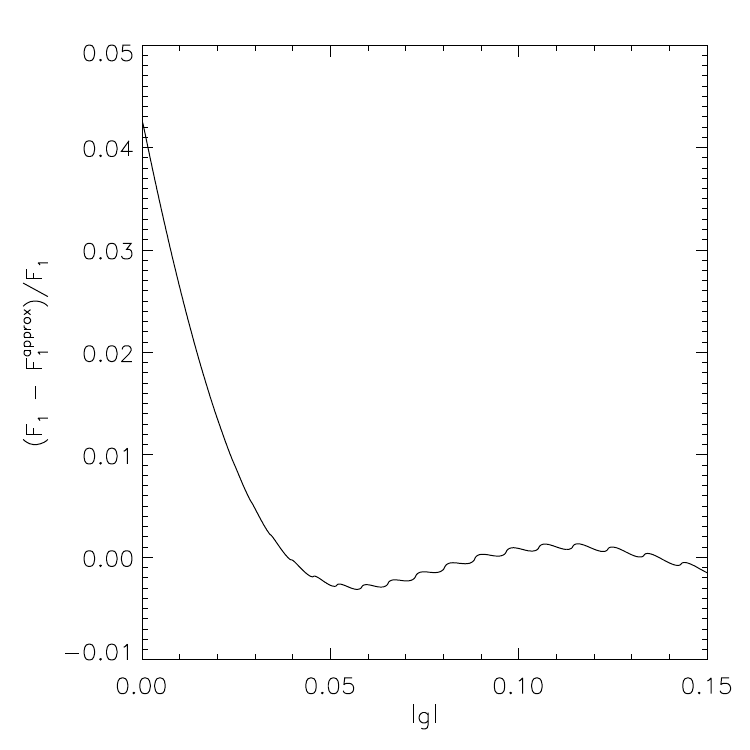}
\caption{The fractional difference between the full $F_1\left(\left|\bm{g}\right|\right)$ function and the 3rd order approximation used to estimate the shear for the simulations. We see that the fractional difference is less than $5\%$ for the range of $\left|\bm{g}\right|$ considered.}
\label{fig:frac_F_logn}
\end{figure} 

When performing the reconstruction using the position angle only method we assume that 
$f\left(\left|\bm{\epsilon}^{\mathrm{int}}\right|\right)$ has been estimated from a large sample of high signal to 
noise galaxy images (see Section \ref{sec:errors_pofe}), such that residual biasing due to an imperfect knowledge of 
the distribution is negligible. To recover the shear estimates we used the corrected form of the 3rd order estimator given in
equation (\ref{eq:3rd_order_mod_unit}); where the best fit parameters are found to be $u=2.423$, $v=4.557$ and $w=-17.465$. Figure \ref{fig:frac_F_logn} shows the fractional difference between the full $F_1\left(\left|\bm{g}\right|\right)$ function and the approximate form used in this analysis. The latter is accurate
to within $5\%$ for all values of $\left|\bm{g}\right|$ in the range considered.

For this set of simulations we assume that the measurement errors 
on the position angles are independent of the true position angles 
and are Gaussian distributed, with zero mean and 
standard deviation $\sigma_{\alpha}=15^{\circ}$. We are, thus, ignoring any effects 
which may contribute to the covariance terms in equation (\ref{eq:mean_c_s}) and which 
would arise as a result of the method used to measure the position angles. Since it is 
difficult to identify a level of ellipticity measurement that directly corresponds to 
our choice of $\sigma_{\alpha}$, for the purpose of our simulation we
have assumed a zero measurement error on the ellipticity
measurements. This would obviously not be the case in real data and so
the reader should bear in mind that the precision of the
ellipticity-based reconstruction, relative to that of the position
angle only reconstruction, will be somewhat over-estimated.

If we assume that the shear is small enough so that, for a given ellipticity distribution, we can use a first order approximation of
the $F_1\left(\left|\bm{g}\right|\right)$ function, then one can
easily obtain a first order approximation to the error on the 
angle only estimator. In such a case the first order corrected estimator is
found to be
\begin{equation}\label{eq:1st_order_gen_est}
\hat{\bm{g}}=\frac{1}{u\beta N}\sum_{i=1}^N\bm{n}^{(i)},
\end{equation}
where $u$ is the first order coefficient of the expansion given in
equation (\ref{eq:F_condition}), and where $\beta$ is the noise bias 
correction term corresponding to a Gaussian measurement error on the 
position angles of $15^{\circ}$. The first order error is therefore
given by
\begin{equation}\label{eq:1st_order_gen_error}
\sigma_{\hat{\bm{g}}}=\frac{1}{u\beta\sqrt{N}}\sigma_{\bm{n}}.
\end{equation}
If we also assume that the shear is small enough that we can
approximate $\sigma_{\bm{n}}^2\approx0.5$ (which is the
maximum value that $\sigma_{\bm{n}}^2$ can take and, from equation (\ref{eq:theor_variance}), is correct to first order in $\left|\bm{g}\right|$), then
equation~(\ref{eq:1st_order_gen_error}) simplifies to
\begin{equation}\label{eq:1st_order_gen_error_simp}
\sigma_{\hat{\bm{g}}}\approx\frac{1}{u\beta\sqrt{2N}}.                          
\end{equation}
From equation (\ref{eq:1st_order_gen_error_simp})
we estimate the error on the shear estimates from 
these simulations to be $\sigma_{\hat{\bm{g}}}\approx0.33/\sqrt{N}$.
The error on the shear estimates when using the standard method 
is $\sigma_{\mathrm{st}}=\sigma_{\epsilon}/\sqrt{N}$, as we have assumed 
zero measurement error on the ellipticities. Therefore, we can 
estimate the ratio of the errors using the two methods to be
\begin{equation}\label{eq:error_est_logn}
\frac{\sigma_{\hat{\bm{g}}}}{\sigma_{\mathrm{st}}}\approx1.3.
\end{equation}

To perform the reconstruction, for each map pixel of side 1 arcmin,
the shear is estimated by assigning a weight to each of the galaxies in
the field so that the shear estimate for each map pixel contains a
contribution from all of the galaxies in the field. For this analysis
we adopted the Gaussian weighting function:
\begin{equation}\label{eq:gauss_weight}
W_k^{(i,j)}=\exp\left({-\frac{\left(\bm{\theta}_{i,j}-\bm{\theta}_k\right)^2}{2\theta_0^2}}\right),
\end{equation}
where $\theta_0$ is the smoothing scale, which, in this case, is taken
to be 1.5 arcmin.

The shear in each pixel in then estimated using both the standard
estimator and the position angle only estimator in order to produce two
shear maps. When using the position angle only approach, the weighting is
applied to the observed unit vectors so that the average unit vector
which describes the average orientation is given by
\begin{equation}\label{eq:gauss_unit_ave}
\left<\bm{n}\right>_{i,j}=\frac{\sum_{k=1}^NW_k^{(i,j)}\bm{n}_k}{\sum_{k=1}W_k^{(i,j)}}.
\end{equation}
In the standard approach the weighting is applied to the galaxy ellipticities.

The convergence field is estimated separately for the two approaches
using the discrete Kaiser-Squires inversion \citep{kaiser93}, which
is given as a convolution of the shear with the kernel
\begin{equation}\label{eq:kappa_kernal}
\mathcal{D}\left(\bm{\theta}\right)=-\frac{1}{\left(\theta_1-i\theta_2\right)^2},
\end{equation}
such that the convergence is estimated as
\begin{equation}\label{eq:discrete_ks93}
\hat{\kappa}\left(\bm{\theta}\right) = \frac{1}{\pi}\sum_{i,j}=\text{Re}\left[\mathcal{D}^{*}\left(\bm{\theta}-\bm{\theta}_{i,j}\right)\hat{\bm{g}}\left(\bm{\theta}_{i,j}\right)\left(1-\hat{\kappa}\left(\bm{\theta}_{i,j}\right)\right)\right],
\end{equation}
which is solved iteratively.

The reconstructed convergence maps are shown in the lower panels of
Fig.~\ref{fig:kappa_sim}. We see that the position angle only approach
successfully recovers the major mass concentrations with a
performance that is qualitatively similar to that of the standard
estimator.

In order to quantify the level of agreement between the two
reconstructions, we have compared the residual map obtained from the
difference between the two reconstructions to the residuals one would
expect to see solely due to noise. To perform the comparison, we
simulated 100 maps containing only galaxy shape noise (and measurement
noise with $\sigma_\alpha = 15^\circ$ in the case of the position
angle analysis). We then repeated the mass reconstructions for each
realization for both the position angle only approach and for the
standard approach. A set of 100 simulated residual maps was then
constructed by taking the difference between the noise-only maps
recovered by the two approaches. 

\begin{figure}
\centering
\includegraphics{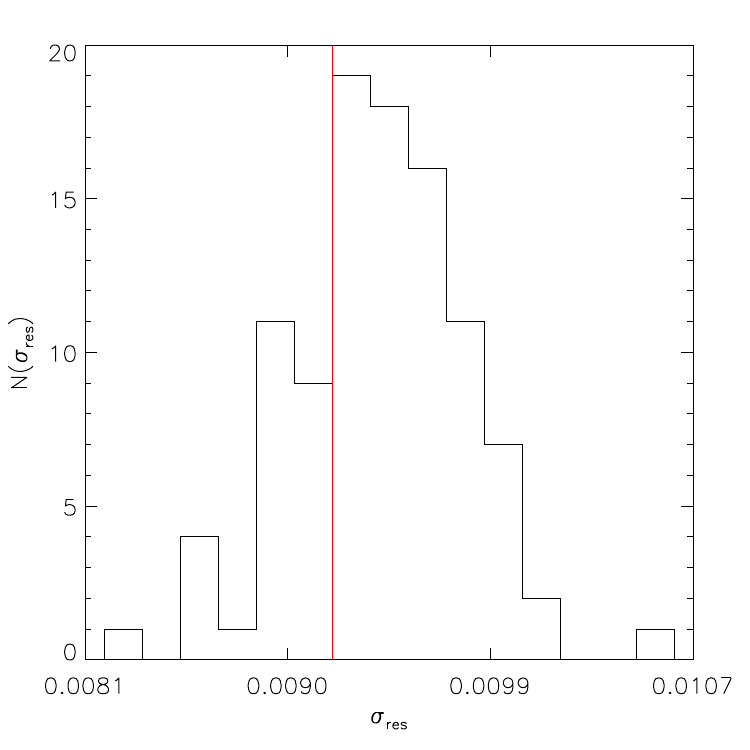}
\caption{The distribution of the r.m.s. residuals,
  $\sigma_{\mathrm{res}}$, obtained from 100 simulated and differenced
  maps containing only noise contributions. The vertical red line
  shows the r.m.s. residual obtained from the difference of the
  reconstructed mass maps  shown in Fig.~\ref{fig:kappa_sim}. This
  plot shows that the difference between the two reconstructed mass
  maps is consistent with noise.}
\label{fig:kappa_res}
\end{figure}

Fig.~\ref{fig:kappa_res} shows a histogram of the r.m.s. residuals,
$\sigma_{\mathrm{res}}$, as measured from the suite of noise-only
difference maps. When calculating the residuals we ignored all pixels
that lay within 5 arcmin of the edge of the reconstructed maps in
order to avoid edge effects. The vertical red line shows the value of
$\sigma_{\mathrm{res}}$ obtained from the difference of the two lower
panels shown in Fig.~\ref{fig:kappa_sim}. Since this is consistent
with having been drawn randomly from the histogram distribution, we
conclude that the two convergence reconstructions are consistent with
each other. The simulated noise maps also provide us with an estimate
of the error on both the standard estimator and the position angle only
estimator for the case of zero shear. The ratio of these errors was
found to be
$\sigma_{\hat{\bm{g}}}/\sigma_{\mathrm{st}}\approx1.3$, which
agrees with equation (\ref{eq:error_est_logn}).

\section{Demonstration on the CFHTL\MakeLowercase{en}S data}
\label{sec:CFHT}
The CFHTLenS \citep{heymans12} has observed four distinct fields, W1
($\sim$63.8$\text{deg}^2$), W2 ($\sim$22.6$\text{ deg}^2$), W3
($\sim$44.2$\text{deg}^2$) and W4 ($\sim$23.3$\text{ deg}^2$),
providing full ellipticity measurements for each detected galaxy in
addition to weights associated with each measurement. As a means of
comparing our position angle only approach with the standard approach using
real data we have reconstructed the mass maps of all four fields using
both techniques.

The weighting provided for each ellipticity measurement has a contribution from the intrinsic shape
dispersion and a measurement error, such that the weighting of the
$i^{\mathrm{th}}$ galaxy is given as \citep{miller13}
\begin{equation}\label{eq:cfht_var}
w_i=\left[\frac{\sigma_i^2\left|\epsilon_{\mathrm{max}}^{\mathrm{int}}\right|^2}{\left|\epsilon_{\mathrm{max}}^{\mathrm{int}}\right|^2-2\sigma_i^2}+\sigma_{\epsilon}^2\right]^{-1},
\end{equation}
where $\sigma_i$ is the measurement error associated with the
ellipticity of the $i^{\mathrm{th}}$ galaxy. The 1D dispersion in the
intrinsic galaxy shapes is taken to be $\sigma_{\epsilon}=0.2539$
\citep{miller13}. There are also two calibration values for each
galaxy, $(1+m_i)$ and $c_{2,i}$, which are, respectively,
multiplicative and additive corrections. The $c_{2,i}$ correction is
deducted from the $\epsilon_2^{\mathrm{obs}}$ component of the
ellipticity and the $(1+m_i)$ correction is applied to the average
ellipticity in a given pixel, such that the standard shear estimator
for each pixel is found to be \citep{waerbeke13}
\begin{equation}\label{eq:standard_cfht_est}
\hat{\bm{g}}=\frac{\sum_{i=1}^Nw_i\left(\bm{\epsilon}_i^{\mathrm{obs}}-\bm{c}_i\right)}{\sum_{i=1}^Nw_i\left(1+m_i\right)},
\end{equation}
where $\bm{c}_i=\left(0,c_{2,i}\right)$.  Following the same procedure
as in Section \ref{sec:sims}, we applied the Gaussian weight given in
equation (\ref{eq:gauss_weight}) to each galaxy so that the
estimated shear in each pixel contains a contribution from all of the
galaxies in a particular field. When considering the CFHTLenS data,
this weighting process allows for a shear to be obtained for the
regions that are masked. A cut-off was applied, such that, if the
contribution to a given pixel has a maximum weighting of less than
0.5, then this pixel is ignored during the reconstruction. 

In order to perform the position angle only analysis we have chosen to
reconstruct the orientation of each galaxy by using equation
(\ref{eq:obs_ellip_comp_angle}) to obtain the relation
\begin{equation}\label{eq:angle_obs}
\alpha=\frac{1}{2}\tan^{-1}\left(\frac{\epsilon_2^{\mathrm{obs}}-c_{2,i}}{\epsilon_1^{\mathrm{obs}}}\right).
\end{equation}
The multiplicative calibration factor is identical for both components
of the ellipticity and so cancels out during the calculation of the
position angles. However, the additive correction must be applied. We
note that position angles reconstructed from the ellipticity
measurements in this way would presumably retain many of the systematics that
might already be present in the ellipticity measurements. In future
work we intend to assess the potential advantages of a position angle
approach to weak lensing in a more comprehensive manner by measuring
the galaxy position angles directly from the imaging data using, for example, the method 
discussed in Section \ref{sec:pos_angles}. However,
our current goal is to demonstrate the feasibility of our proposed
technique for which reconstructing the position angles from the
already carefully measured and calibrated ellipticity estimates,
suffices. 

To implement the position angle only approach we also require an
estimate of the intrinsic ellipticity distribution. To obtain this, we
assumed that all of the galaxies in the CFHTLenS data are disc
dominated -- \cite{miller13} state that this accounts for approximately
90\% of the galaxy population in the survey. In this case, and
assuming that the shear signal is small, the functional form of the
intrinsic ellipticity distribution can be approximated using the prior
distribution \citep{miller13}
\begin{equation}\label{eq:CFHT_dist}
f\left(\left|\bm{\epsilon}^{\mathrm{int}}\right|\right)=\frac{K\left(1-\exp\left(\frac{\left|\bm{\epsilon}^{\mathrm{int}}\right|-\left|\bm{\epsilon}_{\mathrm{max}}^{\mathrm{int}}\right|}{\sigma_{\epsilon}}\right)\right)}{\left(1+\left|\bm{\epsilon}^{\mathrm{int}}\right|\right)\left(\left|\bm{\epsilon}^{\mathrm{int}}\right|^2+\epsilon_0^2\right)^{\frac{1}{2}}},
\end{equation}
where $K$ is a constant which was determined numerically to normalize
the probability. The maximum ellipticity cut-off used was
$\left|\bm{\epsilon}_{\mathrm{max}}^{\mathrm{int}}\right|=0.804$,
which arises primarily from the finite thickness of the galaxy
discs. The dispersion was $\sigma_{\epsilon}=0.2539$ and the
``circularity'' parameter was $\epsilon_0=0.0256$. We tabulated the $F_1\left(\left|\bm{g}\right|\right)$ 
function corresponding to equation (\ref{eq:CFHT_dist}) and inverted the function, using equation (\ref{eq:mod_shear_corr}), 
to obtain an estimate of $\left|\bm{g}\right|$.

Finally, to complete the position angle only estimator, we require a
correction term in order to remove the noise bias associated with the
measurement errors on the galaxy position angles. As described above,
the position angle estimates were derived from the ellipticity
measurements. For our current implementation, the position angle
errors will therefore arise due to the propagation of the ellipticity
errors through equation (\ref{eq:angle_obs}).

Assuming that the measurement errors are independent of the galaxy orientation, 
we obtained the noise bias correction term using the procedure
outlined in Appendix A. We began by assuming that the measurement
errors on the ellipticities are Gaussian distributed
\citep{miller13}. In this case, the probability distribution for the
estimated ellipticity, $\hat{\bm{\epsilon}}$, of the $i^{\rm th}$
galaxy is
\begin{equation}\label{eq:obs_ellip_dist}
f^{(i)}\left(\hat{\bm{\epsilon}}\right)=K\exp\left(-\frac{\left|\hat{\bm{\epsilon}}-\bm{\epsilon}_i^{\mathrm{true}}\right|^2}{2\sigma_i^2}\right),
\end{equation}
where $\bm{\epsilon}_i^{\rm true}$ is the true ellipticity of the galaxy
and the measurement error, $\sigma_i$, can be calculated using equation (\ref{eq:cfht_var}). To arrive at the probability distribution
for the estimated position angle ($\hat{\alpha}$), we must now marginalize over
$\left|\hat{\bm{\epsilon}}\right|$:
\begin{equation}\label{eq:obs_ellip_marg}
f^{(i)}\left(\hat{\alpha}\right)=\int_0^1 \!
f^{(i)}\left(\hat{\bm{\epsilon}}\right) |\hat{\bm{\epsilon}}| \, d|\hat{\bm{\epsilon}}|.
\end{equation}
An exact implementation of equation~(\ref{eq:obs_ellip_marg}),
requires knowledge of the true value of the ellipticity of each
galaxy, $\bm{\epsilon}_i^{\rm true}$, which is obviously not known. In
order to calculate the noise bias correction term, we have therefore
used equation~(\ref{eq:obs_ellip_marg}) with the approximation that
$\bm{\epsilon}_i^{\rm true} \approx \bm{\epsilon}_i^{\rm obs}$, where
$\bm{\epsilon}_i^{\rm obs}$ is the observed ellipticity. While
this will obviously not hold for each individual galaxy, we expect
that after averaging over all of the galaxies in the survey, the
derived mean correction term will be approximately correct. 

The marginalized distribution returned by
equation~(\ref{eq:obs_ellip_marg}) will be symmetrically distributed about
the observed position angle value. When this distribution is shifted,
such that the mean value is at zero, we recover the distribution of
the measurement error on the position angle, $\delta\alpha$. For each
galaxy we then found the value of
$\left<\cos\left(2\delta\alpha\right)\right>_i$. Doing this for
all galaxies in a particular field of the CFHTLenS data, the final correction term is simply
the mean of all of the values of
$\left<\cos\left(2\delta\alpha\right)\right>_i$. The bias correction was found independently 
for each field. All four corrections were found to be $\beta\approx0.8$ which corresponds to an equivalent Gaussian
measurement error of $\sigma_{\alpha}\approx19^{\circ}$. It is conceivable
that this error could be reduced if, in future surveys, measurements
of the position angles are obtained directly from the imaging data as discussed in Section \ref{sec:pos_angles}.

\begin{figure*}
\begin{minipage}{6in}
\centering
\includegraphics{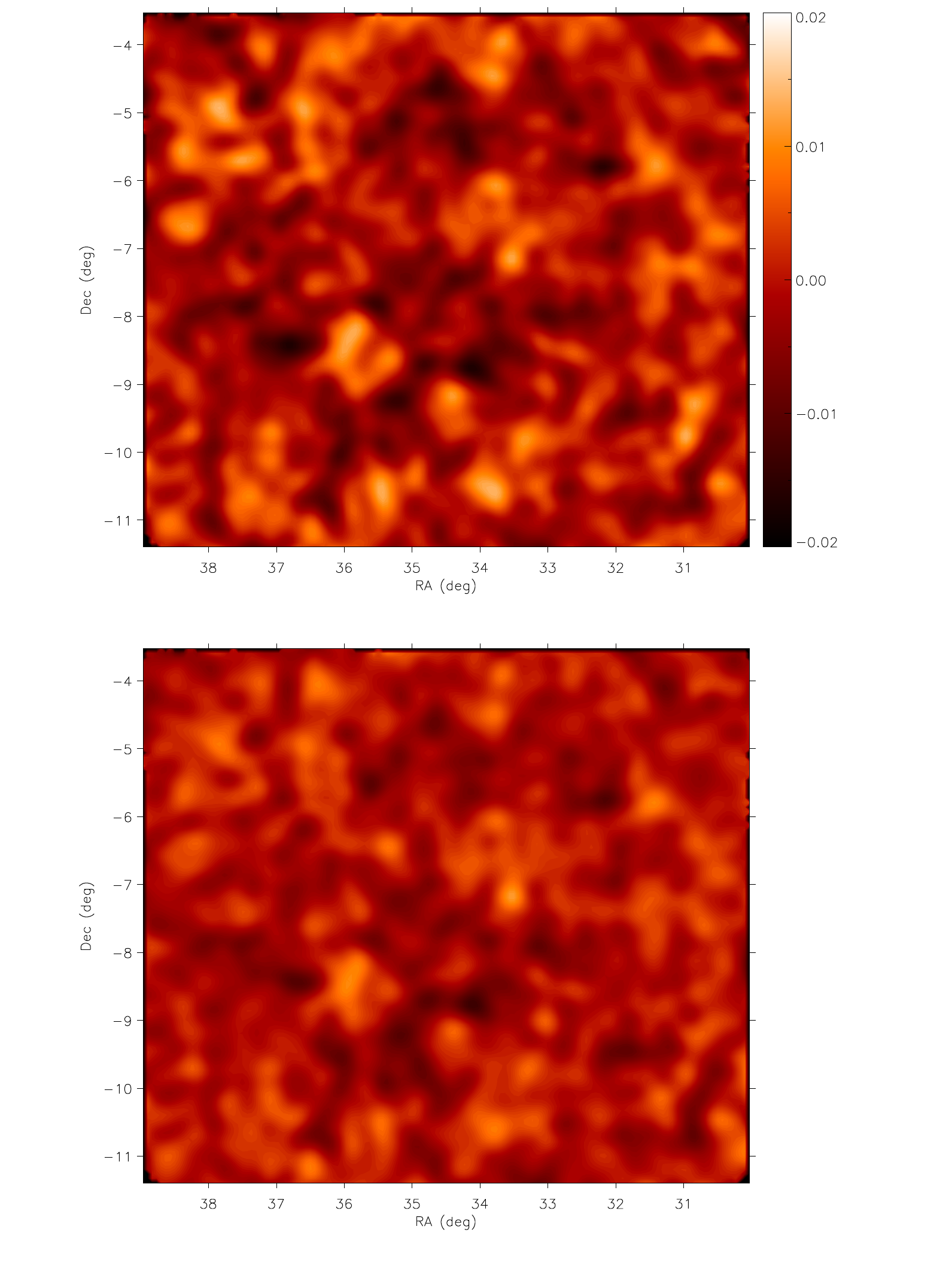}
\caption{Mass reconstructions for the W1 field of the CFHTLenS. The
  top panel shows the reconstruction obtained using the standard
  method and the bottom panel is performed using the position angle only
  approach. The smoothing scale for these reconstructions is 8.9
  arcmin. The colour bars indicate the scale of the convergence fields.}
\label{fig:kappa_cfht}
\end{minipage}
\end{figure*}

\begin{figure}
\centering
\includegraphics{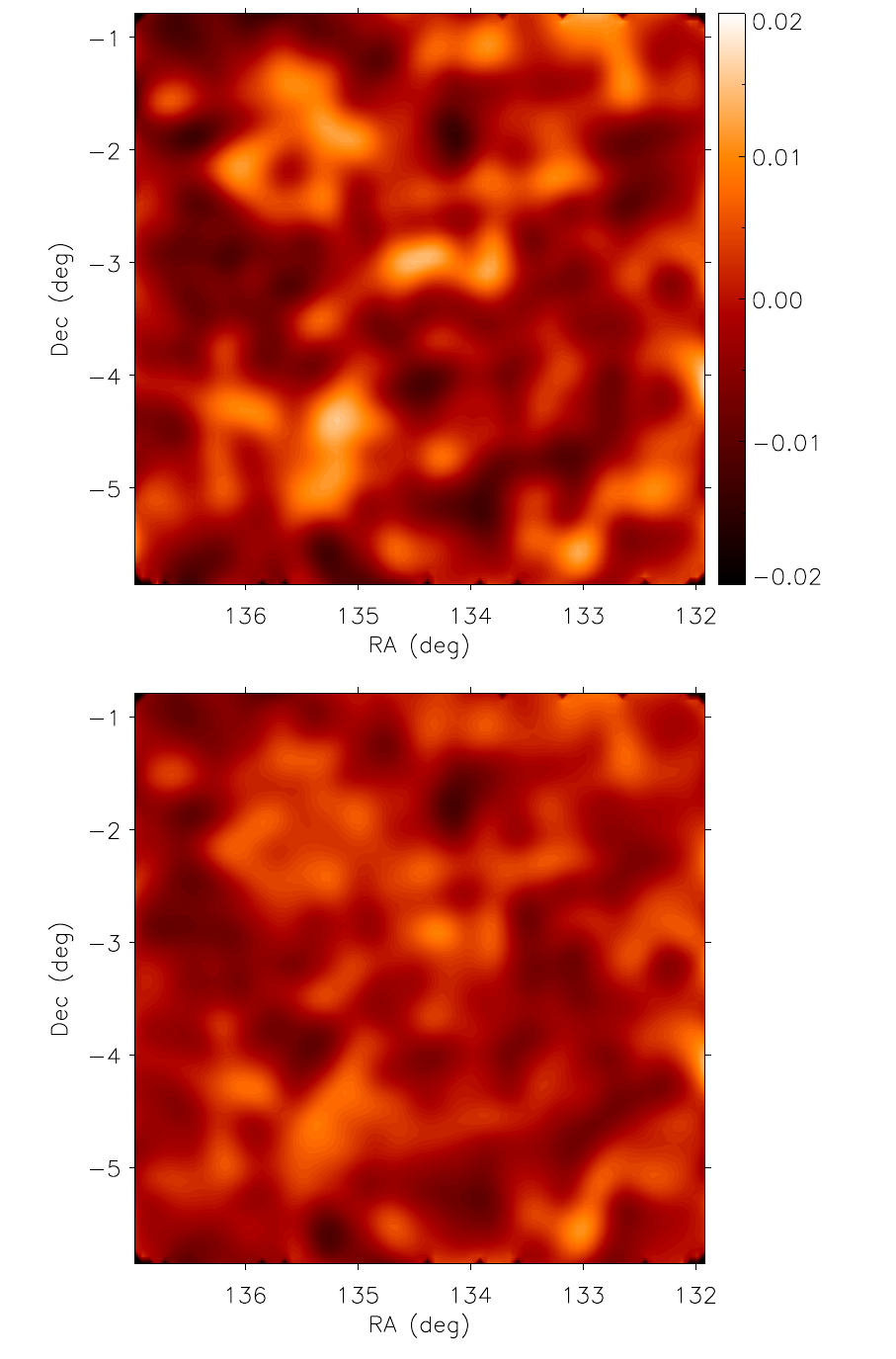}
\caption{Mass reconstructions for the W2 field of the CFHTLenS. The
  top panel shows the reconstruction obtained using the standard
  method and the bottom panel is performed using the position angle only
  approach. The smoothing scale for these reconstructions is 8.9
  arcmin. The colour bars indicate the scale of the convergence fields.}
\label{fig:kappa_cfht2}
\end{figure}

\begin{figure*}
\begin{minipage}{6in}
\centering
\includegraphics{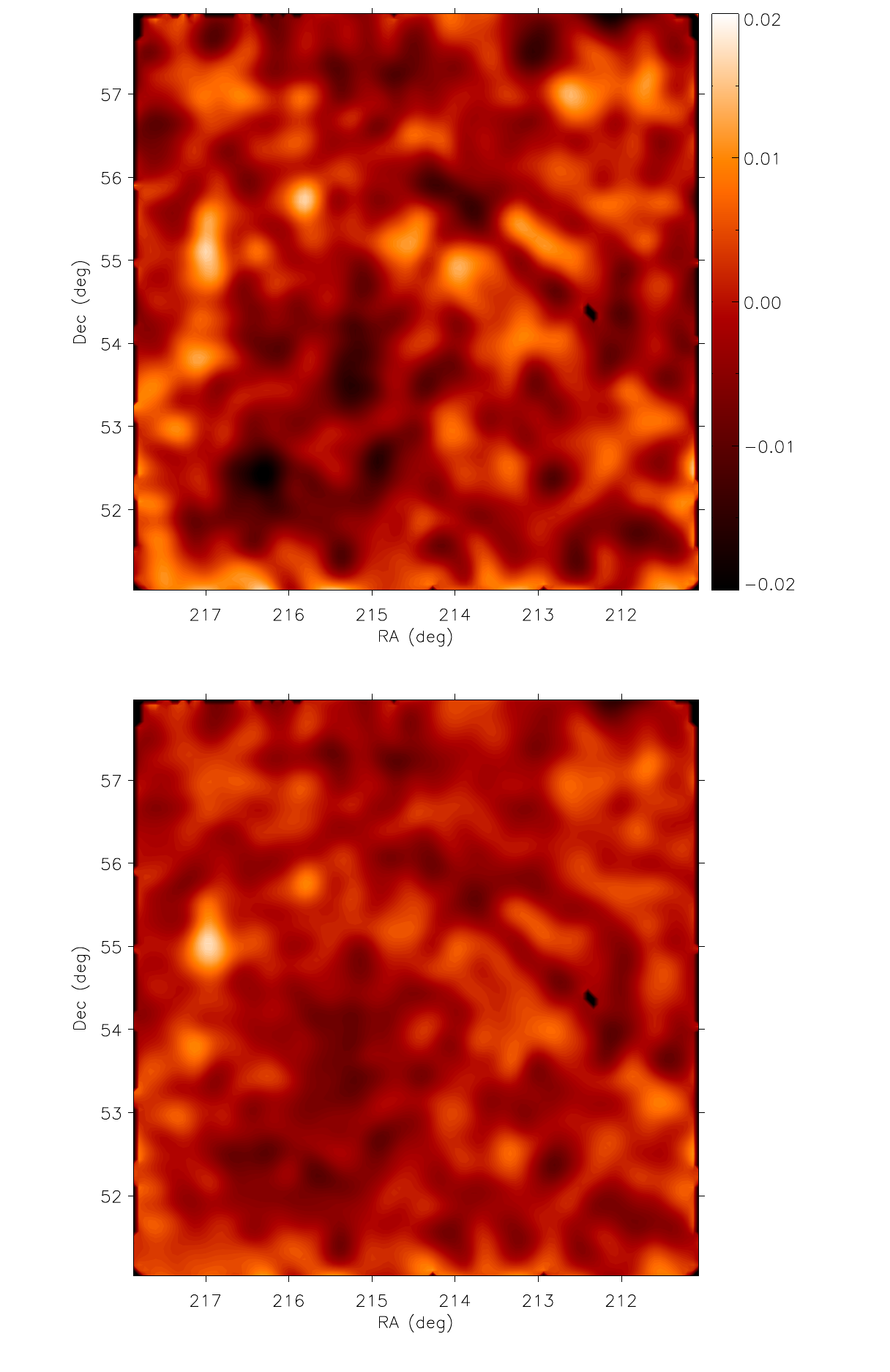}
\caption{Mass reconstructions for the W3 field of the CFHTLenS. The
  top panel shows the reconstruction obtained using the standard
  method and the bottom panel is performed using the position angle only
  approach. The smoothing scale for these reconstructions is 8.9
  arcmin. The colour bars indicate the scale of the convergence fields.}
\label{fig:kappa_cfht3}
\end{minipage}
\end{figure*}

\begin{figure}
\centering
\includegraphics{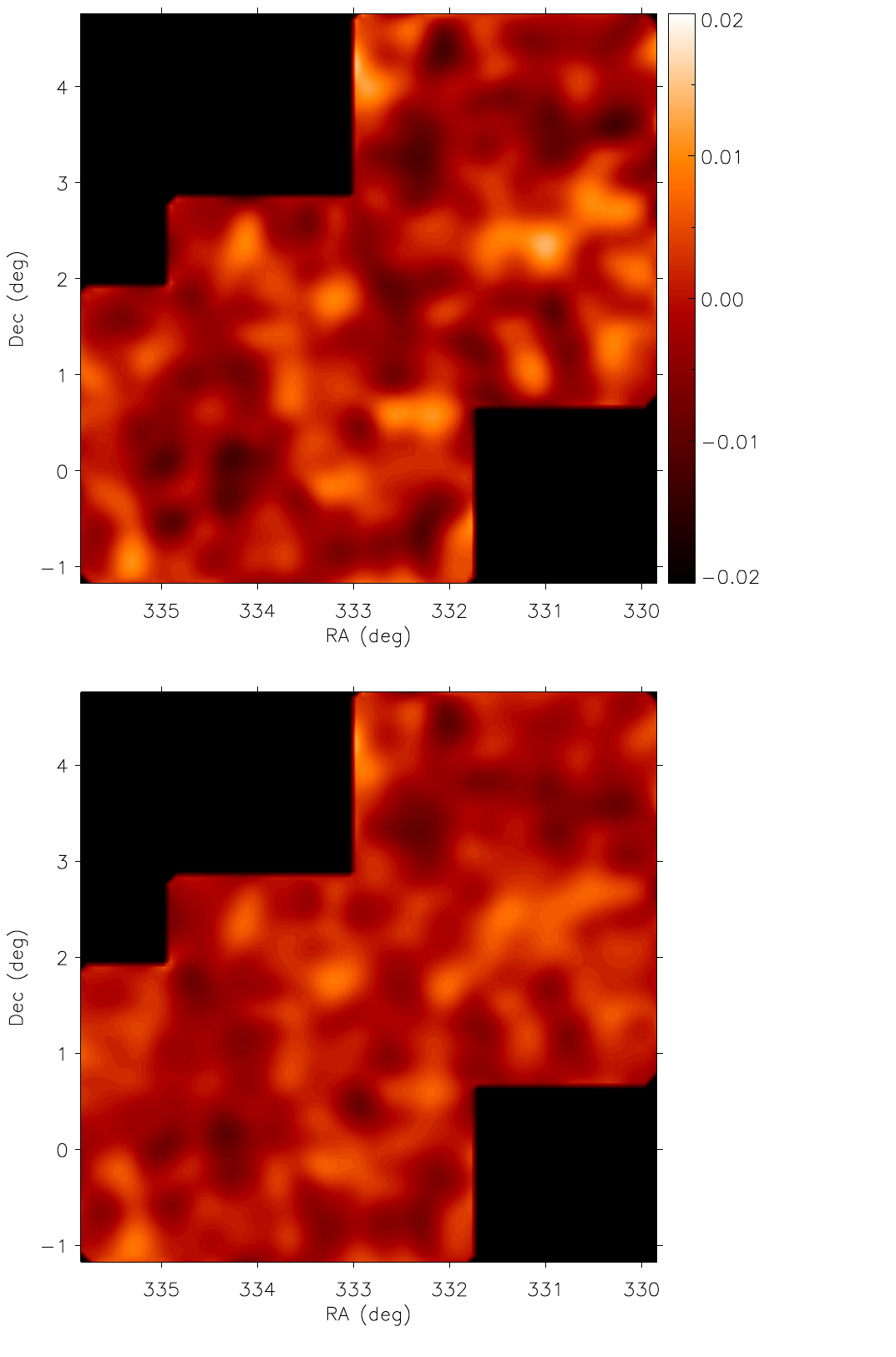}
\caption{Mass reconstructions for the W4 field of the CFHTLenS. The
  top panel shows the reconstruction obtained using the standard
  method and the bottom panel is performed using the position angle only
  approach. The smoothing scale for these reconstructions is 8.9
  arcmin. The colour bars indicate the scale of the convergence fields.}
\label{fig:kappa_cfht4}
\end{figure}

Using the angle only method,
and following the approach adopted in Section \ref{sec:sims}, we
constructed mass maps for each of the four fields in the CFHTLenS. The
pixel size used to reconstruct the maps was 4 arcmin, with a smoothing
scale of 8.9 arcmin, which is the same as used in \cite{waerbeke13}.
The resulting maps are shown in
Figs.~\ref{fig:kappa_cfht}--\ref{fig:kappa_cfht4}, where we 
also present maps reconstructed using the standard (full ellipticity)
approach for visual comparison. One can immediately see the
qualitative agreement in the maps reconstructed using the two
methods -- the position angle only approach recovers mass
concentrations at the same locations as the standard estimator and on
a similar scale.

\begin{figure*}
\begin{minipage}{6in}
\centering
\includegraphics{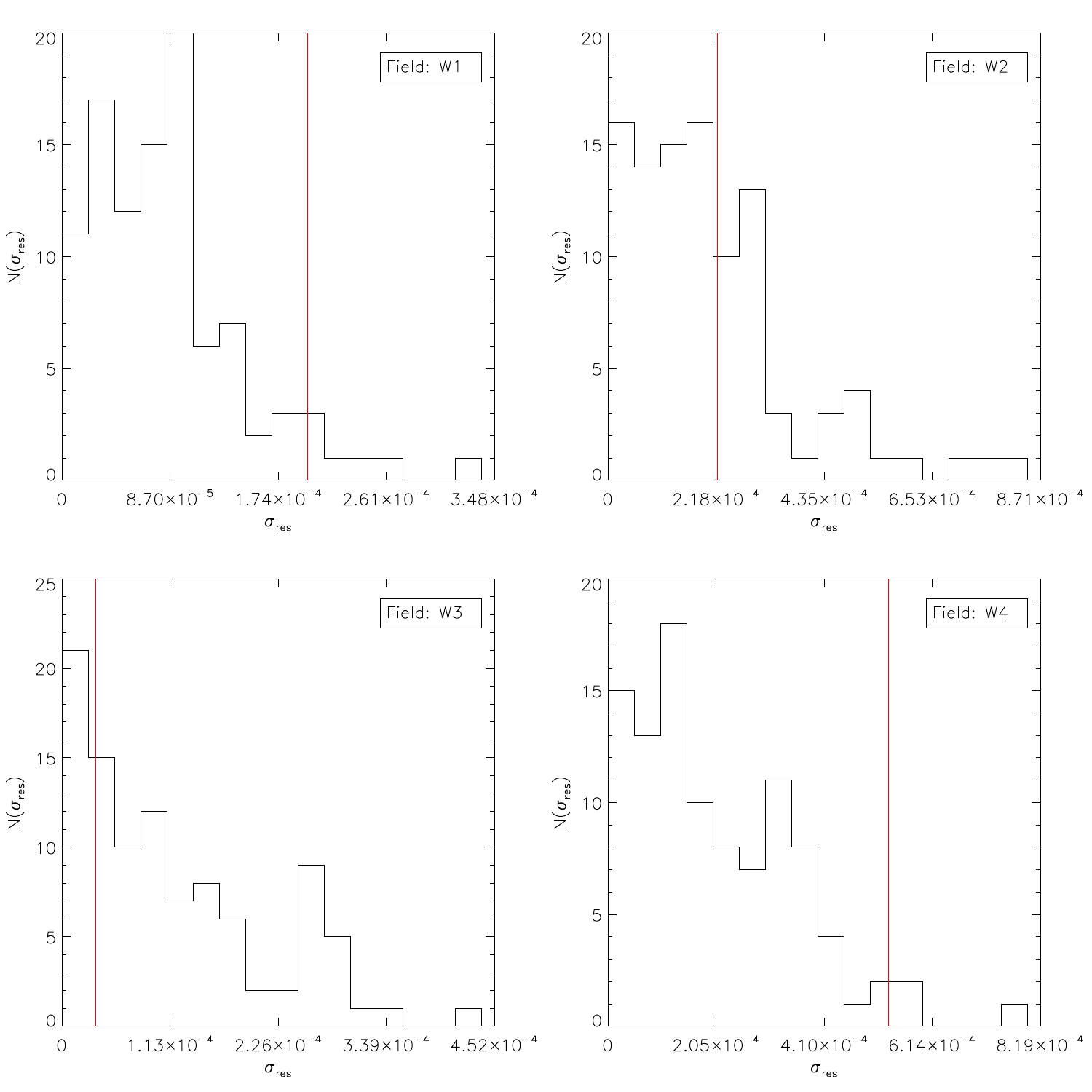}
\caption{The distribution of the r.m.s residuals,
  $\sigma_{\mathrm{res}}$, obtained from 100 pairs of reconstructed
  maps containing only noise for each of the four fields of the
  CFHTLenS. The vertical red lines show the values of
  $\sigma_{\mathrm{res}}$ obtained from the difference between the
  reconstructed maps shown in
  Figs.~\ref{fig:kappa_cfht}--\ref{fig:kappa_cfht4}. These results
  indicate that the difference between the two sets of reconstructed
  mass maps is consistent with noise.}
\label{fig:kappa_res_cfht}
\end{minipage}
\end{figure*}

Following the approach described in Section \ref{sec:sims} we have
attempted to quantify the level of agreement between the two sets of
mass reconstructions by making use of simulations including only the
effects of shape and measurement noise. For the case of the CFHTLenS
data we created noise realizations by assigning a random orientation
to every galaxy in the dataset. We performed two mass reconstructions
for each realization -- one using the standard method and one using the
position angle only method to estimate the shear. We produced a
histogram of the r.m.s. residuals as measured from the difference
between the reconstructed maps. The results are shown in
Fig.~\ref{fig:kappa_res_cfht} (black curves). When calculating the
residuals we ignored all pixels in the mass maps that lie within
$\sim$30 arcmin of a masked region in order to reduce edge
effects. The vertical red lines show the r.m.s. residuals obtained from
difference maps constructed from the mass reconstructions shown in
Figs.~\ref{fig:kappa_cfht}--\ref{fig:kappa_cfht4}. For each of the
four fields, the residual maps obtained by differencing the mass
reconstructions are shown to be consistent with the corresponding
difference maps from the simulations containing only noise. These
results suggest that systematic differences between the two shear
estimation techniques are sub-dominant to the noise in the
reconstructions.  

\section{Conclusions}
\label{sec:conclusions}
Building on the work of \cite{schneider95} we have demonstrated a
method of performing a weak lensing analysis using only the position
angle measurements for a set of galaxies. By using the probability
distribution for the intrinsic ellipticities of the galaxies one can
express the mean of the trigonometric functions in terms of the
underlying shear unit vector and a function that depends on the
modulus of the shear. Obtaining an estimate of the shear components is
then possible by means of inverting this relationship. It has been 
shown that the bias introduced by position angle measurement errors 
can be reduced to negligible levels by the introduction of a correction 
term, which can be calculated numerically for a general error distribution.

The method has been successfully demonstrated using both simulations
and using the data from the CFHTLenS. Upon comparing
the residuals of the mass maps constructed using both the standard
method and the position angle only method, we have demonstrated that
the difference between the two approaches is consistent with
noise. This demonstration was performed using simulations where the
intrinsic ellipticity distribution was assumed to be a log-normal
distribution, and using the data from the CFHTLenS, where we used the
best-fit intrinsic ellipticity distribution for the disc dominated
galaxies \citep{miller13}. 

For the demonstration of our method on the CFHTLenS data, we derived
the position angle estimates from the ellipticity measurements
provided with the CFHTLenS data release. In order to fully exploit the
potential advantages of position angle based weak lensing analyses,
we have introduced a method of measuring the position angles of the galaxies 
directly from the imaging data. This method will be developed in future
work with the goal of reducing systematics and complementing
parallel weak lensing analyses based on the full ellipticity
information.  

In the absence of direct position angle measurements we find that when
we calculate the position angles from the ellipticities provided by
CFHTLenS, the multiplicative bias cancels out. Our position angle
only shear estimates will therefore only be sensitive to additive
biases. However, for the position angle only method to be successful, it is
vital that the correct form of the intrinsic ellipticity distribution
is used for any particular survey. If an incorrect form is used, then
the form of the $F_1\left(\left|\bm{g}\right|\right)$ function
will be incorrect and this will itself lead to mis-calibrated
shear estimates. It has been shown that a realistic sample size of high resolution 
galaxy images can be used to obtain an estimate of the intrinsic ellipticity distribution 
such that residual biases in the shear estimates, resulting from an incorrect distribution, 
are negligible.

In future surveys where the polarization information associated with
radio emission will be available, such as those conducted by
e-MERLIN\footnote{http://www.e-merlin.ac.uk/legacy/projects/superclass.html},
the methods described in this paper could be used to provide
information about the intrinsic orientation of the galaxies
\citep{brown11a}, and thereby provide a method of mitigating intrinsic
alignment contamination in future weak lensing surveys.

\renewcommand{\theequation}{A-\arabic{equation}}
\renewcommand{\thefigure}{A-\arabic{figure}}
\setcounter{equation}{0}  
\setcounter{figure}{0}
\section*{Appendix A: Including a variable $\sigma_{\alpha}$}
\label{var_sig}
Assuming that the measurement errors are independent of the position angles 
and symmetrically distributed about zero, we derive the bias correction 
for a general distribution of measurement errors.

We begin by writing the measured position angle as
$\hat{\alpha}=\alpha+\delta\alpha$, where $\delta\alpha$ is a random
measurement error. Assuming that this error is independent of
$\alpha$, we can write the mean unit vector components as
\begin{align}\label{eq:est_trig_bias1_append}
\left<\cos\left(2\hat{\alpha}\right)\right>&=\left<\cos\left(2\alpha\right)\right>\left<\cos\left(2\delta\alpha\right)\right>-\left<\sin\left(2\alpha\right)\right>\left<\sin\left(2\delta\alpha\right)\right>,\nonumber\\
\left<\sin\left(2\hat{\alpha}\right)\right>&=\left<\sin\left(2\alpha\right)\right>\left<\cos\left(2\delta\alpha\right)\right>+\left<\cos\left(2\alpha\right)\right>\left<\sin\left(2\delta\alpha\right)\right>.
\end{align}
Making the further assumption that $\delta\alpha$ is distributed
symmetrically about zero, we obtain a relation between the mean measured 
unit vector components and the corrected unit vector components
\begin{equation}\label{correct_trig_gen}
\left<\bm{n}\right>=\left<\bm{n}\right>^{\mathrm{corrected}}\left<\cos\left(2\delta\alpha\right)\right>.
\end{equation}
In practice the distribution of $\delta\alpha$ may be different for
each galaxy, and, therefore, one would need to determine the
multiplicative factor for the $i^{\mathrm{th}}$ galaxy,
$\left<\cos\left(2\delta\alpha\right)\right>_i$. If we define the overall
correction term, $\beta$, where
\begin{equation}\label{eq:correct_term}
\left<\bm{n}\right>^{\mathrm{corrected}}=\frac{\left<\bm{n}\right>}{\beta},
\end{equation}
we can determine $\beta$ by finding the mean of the individual multiplicative factors, $\left<\cos\left(2\delta\alpha\right)\right>_i$, such that
\begin{equation}\label{eq:geberal_beta}
\beta=\frac{1}{N}\sum_{i=1}^N\left<\cos\left(2\delta\right)\right>_i.
\end{equation}

As an example, let us assume that the measurement errors are Gaussian
distributed and that the variance on the error distribution for the
$i^{\mathrm{th}}$ galaxy is $\sigma_{\alpha,i}^2$, then the mean unit 
vector becomes
\begin{align}\label{eq:mean_trig_var_sig}
\left<\bm{n}\right>&=\left<\bm{n}\right>^{\mathrm{corrected}}\left<\cos\left(2\delta\alpha\right)\right>\nonumber\\
&=\left<\bm{n}\right>^{\mathrm{corrected}}\int\mathrm{d}\sigma_{\alpha}f\left(\sigma_{\alpha}\right)\exp\left(-2\sigma_{\alpha}^2\right),
\end{align}
where $f\left(\sigma_{\alpha}\right)$ is the probability density
function of $\sigma_{\alpha}$. If we now make the further assumption
that $\sigma_{\alpha}$ is Gaussian distributed about the mean error,
$\bar{\sigma}_{\alpha}$, with a variance of
$\sigma_{\sigma_{\alpha}}^2$, it can be shown that the corrected mean
unit vector becomes
\begin{equation}\label{eq:mean_trig_gauss_err}
\left<\bm{n}\right>^{\mathrm{corrected}}=\left<\bm{n}\right>\exp\left(\frac{2\bar{\sigma}_{\alpha}^2}{1+4\sigma_{\sigma_{\alpha}}^2}\right)\sqrt{1+4\sigma_{\sigma_{\alpha}}^2},
\end{equation}
such that the correction term is
\begin{equation}\label{eq:example_correct}
\beta=\frac{\exp\left(\frac{-2\bar{\sigma}_{\alpha}^2}{1+4\sigma_{\sigma_{\alpha}}^2}\right)}{\sqrt{1+4\sigma_{\sigma_{\alpha}}^2}}.
\end{equation}
The correction due to a distribution in $\sigma_{\alpha}$ can,
therefore, be attributed to an effective correction. This indicates
that the distribution in errors can be viewed as a single Gaussian
distribution, which for this case, has a variance of
\begin{equation}\label{eq:effective_variance}
\sigma_{\alpha}^2=\frac{\bar{\sigma}^2}{1+4\sigma_{\sigma_{\alpha}}^2}+\frac{1}{4}\ln\left(1+4\sigma_{\sigma_{\alpha}}^2\right).
\end{equation}

\renewcommand{\theequation}{B-\arabic{equation}}
\renewcommand{\thefigure}{B-\arabic{figure}}
\setcounter{equation}{0}  
\setcounter{figure}{0}
\section*{Appendix B: Deriving the 3rd order estimator}
\label{3rd_deriv}
Here we construct a 3rd order shear estimator for the case of a general
intrinsic ellipticity distribution. In order to obtain a direct estimator in the general case, one can numerically obtain the 
$F_1\left(\left|\bm{g}\right|\right)$ function that corresponds
to a given intrinsic distribution, 
$f\left(\left|\bm{\epsilon}^{\mathrm{int}}\right|\right)$, by using
equation (\ref{eq:general_F}). Assuming that the shear is much smaller
than the dispersion in the intrinsic ellipticities, we can expand the
$F_1\left(\left|\bm{g}\right|\right)$ function in powers of
$\left|\bm{g}\right|$. For a zero shear signal there will be no
preferred position angle so that
$F_1\left(\left|\bm{g}\right|\right)\rightarrow0$ as
$\left|\bm{g}\right|\rightarrow0$. We can, therefore, write an
approximate form of the $F_1\left(\left|\bm{g}\right|\right)$
function as
\begin{equation}\label{eq:F_expansion}
F_1\left(\left|\bm{g}\right|\right)\approx u\left|\bm{g}\right|+v\left|\bm{g}\right|^2+w\left|\bm{g}\right|^3,
\end{equation}
for some $u,v,w$. We can fit the approximate form of the
$F_1\left(\left|\bm{g}\right|\right)$ function to a numerically
determined function.

From the measured position angles in a given pixel, we can estimate 
the value of $F_1\left(\left|\bm{g}\right|\right)$, such that
\begin{equation}\label{eq:F_est}
\hat{F}_1=\sqrt{\left[\frac{1}{N}\sum_{i=1}^N\cos\left(2\alpha^{(i)}\right)\right]^2+\left[\frac{1}{N}\sum_{i=1}^N\sin\left(2\alpha^{(i)}\right)\right]^2}.
\end{equation}
By equating this expression with the right hand side of equation
(\ref{eq:F_expansion}) we can obtain an estimate of
$\left|\bm{g}\right|$ which satisfies the condition
\begin{equation}\label{eq:F_condition}
u\left|\hat{\bm{g}}\right|+v\left|\hat{\bm{g}}\right|^2+w\left|\hat{\bm{g}}\right|^3-\hat{F_1}=0.
\end{equation}
Solutions of this equation are obtained from the general solution for
the roots of a 3rd order polynomial, which is
\begin{align}\label{eq:3rd_order_sols}
\left|\hat{\bm{g}}\right|=&\left\{q+\left[q^2+\left(r-p^2\right)^3\right]^{\frac{1}{2}}\right\}^{\frac{1}{3}}+\nonumber\\
&\left\{q-\left[q^2+\left(r-p^2\right)^3\right]^{\frac{1}{2}}\right\}^{\frac{1}{3}}+p,
\end{align}
where
\begin{align}\label{eq:3rd_order_subs}
p=&-\frac{v}{3w},\nonumber\\
q=&\frac{\hat{F}_1}{2w}+\frac{uv}{6w^2}-\frac{v^3}{27w^3},\nonumber\\
r=&\frac{u}{3w}.
\end{align}
Let us now consider the case where $q^2+\left(r-p^2\right)^3<0$. In
this case we can rewrite equation (\ref{eq:3rd_order_sols}) as
\begin{align}\label{eq:3rd_order_lt}
\left|\hat{\bm{g}}\right|=&\left\{q+i\left[\left(p^2-r\right)^3-q^2\right]^{\frac{1}{2}}\right\}^{\frac{1}{3}}+\nonumber\\
&\left\{q-i\left[\left(p^2-r\right)^3-q^2\right]^{\frac{1}{2}}\right\}^{\frac{1}{3}}+p.
\end{align}
However, $\left|\bm{g}\right|$ must be a real solution of
equation (\ref{eq:F_condition}), therefore we can immediately assume
the form of the solution to be
\begin{equation}\label{eq:3rd_order_kest}
\left|\hat{\bm{g}}\right|=2B_0^{\frac{1}{6}}\cos\left[\frac{1}{3}\tan^{-1}\left(\frac{\sqrt{B_1}}{A}\right)+\frac{2n\pi}{3}\right]-\frac{v}{3w},
\end{equation}
where $n=-1,0,1$ and
\begin{align}\label{eq:A;B}
A=&q,\nonumber\\
B_k=&\left(p^2-r\right)^3-kq^2.
\end{align}
The specific choice of $n$ is dependent on the form of the
$F_1\left(\left|\bm{g}\right|\right)$ function, but it will
always be the value of $n$ which minimizes the absolute value of the
cosine term in equation (\ref{eq:3rd_order_kest}). In all of the
simulations that we have conducted we find that $n=-1$.

Let us now examine the case where $q^2+\left(r-p^2\right)^3\ge0$. In
such a case $\left|\hat{\bm{g}}\right|$ can be obtained directly
from equation (\ref{eq:3rd_order_sols}), that is
\begin{equation}\label{eq:3rd_order_ge}
\left|\hat{\bm{g}}\right|=\left(\sqrt{-B_1}+A\right)^{\frac{1}{3}}-\left(\sqrt{-B_1}-A\right)^{\frac{1}{3}}-\frac{v}{3w},
\end{equation}
which can have only one real solution.

An estimate for the orientation of the shear is obtained from equation
(\ref{eq:orient_shear}), such that
\begin{equation}\label{eq:angle_est}
2\alpha_0=\tan^{-1}\left(\frac{\sum_{i=1}^N\sin\left(2\alpha^{(i)}\right)}{\sum_{i=1}^N\cos\left(2\alpha^{(i)}\right)}\right).
\end{equation}
By taking the cosine and sine of equation (\ref{eq:angle_est}) and
dividing through by $N$, it can be shown that the estimated shear unit
vector $\hat{\bm{n}}_0$ can be written as
\begin{equation}\label{eq:est_shear_unit/N}
\hat{\bm{n}}_0=\frac{1}{\hat{F}_1}\left(
\begin{array}{c}
\frac{1}{N}\sum_{i=1}^N\cos\left(2\alpha^{(i)}\right)\\
\frac{1}{N}\sum_{i=1}^N\sin\left(2\alpha^{(i)}\right)
\end{array}\right).
\end{equation}
The full 3rd order estimator can now be obtained, such that
\begin{equation}\label{eq:3rd_order_mod_unit}
\hat{\bm{g}}=\left|\hat{\bm{g}}\right|\hat{\bm{n}}_0.
\end{equation}

If we assume a measurement error on the position angles which is independent 
of the true position angles and drawn from a distribution which is symmetric about zero, then 
we can correct for the measurement error bias using equation (\ref{eq:est_trig_cor}). It can be shown that, 
to correct the third order estimator, we only need to modify the term $q$ in equation (\ref{eq:3rd_order_subs}), such that
\begin{equation}\label{eq:q_correct}
q=\frac{\hat{F}_1}{2w\beta}+\frac{uv}{6w^2}-\frac{v^3}{27w^3}.
\end{equation}

For a more general error distribution we must use the form of the cosines and sines 
given in equation (\ref{eq:mean_c_s}), which may be determined using an iterative method 
as outlined in Subsections \ref{subsec:debias_angles} and \ref{subsec:compare_KSB_angles}.

\section*{Acknowledgments}
LW and MLB are grateful to the European Research Council for support
through the award of an ERC Starting Independent Researcher Grant
(EC FP7 grant number 280127). MLB also thanks the STFC for the award
of Advanced and Halliday fellowships (grant number ST/I005129/1).

The analysis presented in Section~\ref{sec:CFHT} is based on
observations obtained with MegaPrime/MegaCam, a joint project of CFHT
and CEA/IRFU, at the Canada-France-Hawaii Telescope (CFHT) which is
operated by the National Research Council (NRC) of Canada, the
Institut National des Sciences de l'Univers of the Centre National de
la Recherche Scientifique (CNRS) of France, and the University of
Hawaii. This research used the facilities of the Canadian Astronomy
Data Centre operated by the National Research Council of Canada with
the support of the Canadian Space Agency. CFHTLenS data processing was
made possible thanks to significant computing support from the NSERC
Research Tools and Instruments grant program.

\bibliographystyle{mn2e} \bibliography{ms}

\begin{thebibliography}{31}
\expandafter\ifx\csname natexlab\endcsname\relax\def\natexlab#1{#1}\fi

\bibitem[{{Albrecht} {et~al.}(2006){Albrecht}, {Bernstein}, {Cahn}, {Freedman},
  {Hewitt}, {Hu}, {Huth}, {Kamionkowski}, {Kolb}, {Knox}, {Mather}, {Staggs},
  \& {Suntzeff}}]{albrecht06}
{Albrecht} A., {Bernstein} G., {Cahn} R., {Freedman} W.~L., {Hewitt} J., {Hu}
  W., {Huth} J., {Kamionkowski} M., {Kolb} E.~W., {Knox} L., {Mather} J.~C.,
  {Staggs} S., {Suntzeff} N.~B., 2006, ArXiv Astrophysics e-prints

\bibitem[{{Bacon} {et~al.}(2000){Bacon}, {Refregier}, \& {Ellis}}]{bacon00}
{Bacon} D.~J., {Refregier} A.~R., {Ellis} R.~S., 2000, \mnras, 318, 625

\bibitem[{{Bartelmann} \& {Schneider}(2001)}]{bartelmann01}
{Bartelmann} M., {Schneider} P., 2001, \physrep, 340, 291

\bibitem[{{Bridle} {et~al.}(2010){Bridle}, {Balan}, {Bethge}, {Gentile},
  {Harmeling}, {Heymans}, {Hirsch}, {Hosseini}, {Jarvis}, {Kirk}, {Kitching},
  {Kuijken}, {Lewis}, {Paulin-Henriksson}, {Sch{\"o}lkopf}, {Velander},
  {Voigt}, {Witherick}, {Amara}, {Bernstein}, {Courbin}, {Gill}, {Heavens},
  {Mandelbaum}, {Massey}, {Moghaddam}, {Rassat}, {R{\'e}fr{\'e}gier}, {Rhodes},
  {Schrabback}, {Shawe-Taylor}, {Shmakova}, {van Waerbeke}, \&
  {Wittman}}]{bridle10}
{Bridle} S., {Balan} S.~T., {Bethge} M., {Gentile} M., {Harmeling} S.,
  {Heymans} C., {Hirsch} M., {Hosseini} R., {Jarvis} M., {Kirk} D., {Kitching}
  T., {Kuijken} K., {Lewis} A., {Paulin-Henriksson} S., {Sch{\"o}lkopf} B.,
  {Velander} M., {Voigt} L., {Witherick} D., {Amara} A., {Bernstein} G.,
  {Courbin} F., {Gill} M., {Heavens} A., {Mandelbaum} R., {Massey} R.,
  {Moghaddam} B., {Rassat} A., {R{\'e}fr{\'e}gier} A., {Rhodes} J.,
  {Schrabback} T., {Shawe-Taylor} J., {Shmakova} M., {van Waerbeke} L.,
  {Wittman} D., 2010, \mnras, 405, 2044

\bibitem[{{Bridle} {et~al.}(2002){Bridle}, {Kneib}, {Bardeau}, \&
  {Gull}}]{bridle02}
{Bridle} S.~L., {Kneib} J.-P., {Bardeau} S., {Gull} S.~F., 2002, in The Shapes
  of Galaxies and their Dark Halos, {Natarajan} P., ed., pp. 38--46

\bibitem[{{Brown} \& {Battye}(2011{\natexlab{a}})}]{brown11b}
{Brown} M.~L., {Battye} R.~A., 2011{\natexlab{a}}, \apjl, 735, L23

\bibitem[{{Brown} \& {Battye}(2011{\natexlab{b}})}]{brown11a}
---, 2011{\natexlab{b}}, \mnras, 410, 2057

\bibitem[{{Brown} {et~al.}(2003){Brown}, {Taylor}, {Bacon}, {Gray}, {Dye},
  {Meisenheimer}, \& {Wolf}}]{brown03}
{Brown} M.~L., {Taylor} A.~N., {Bacon} D.~J., {Gray} M.~E., {Dye} S.,
  {Meisenheimer} K., {Wolf} C., 2003, \mnras, 341, 100

\bibitem[{{Fu} {et~al.}(2008){Fu}, {Semboloni}, {Hoekstra}, {Kilbinger}, {van
  Waerbeke}, {Tereno}, {Mellier}, {Heymans}, {Coupon}, {Benabed}, {Benjamin},
  {Bertin}, {Dor{\'e}}, {Hudson}, {Ilbert}, {Maoli}, {Marmo}, {McCracken}, \&
  {M{\'e}nard}}]{fu08}
{Fu} L., {Semboloni} E., {Hoekstra} H., {Kilbinger} M., {van Waerbeke} L.,
  {Tereno} I., {Mellier} Y., {Heymans} C., {Coupon} J., {Benabed} K.,
  {Benjamin} J., {Bertin} E., {Dor{\'e}} O., {Hudson} M.~J., {Ilbert} O.,
  {Maoli} R., {Marmo} C., {McCracken} H.~J., {M{\'e}nard} B., 2008, \aap, 479,
  9

\bibitem[{{Heymans} {et~al.}(2012){Heymans}, {Van Waerbeke}, {Miller}, {Erben},
  {Hildebrandt}, {Hoekstra}, {Kitching}, {Mellier}, {Simon}, {Bonnett},
  {Coupon}, {Fu}, {Harnois D{\'e}raps}, {Hudson}, {Kilbinger}, {Kuijken},
  {Rowe}, {Schrabback}, {Semboloni}, {van Uitert}, {Vafaei}, \&
  {Velander}}]{heymans12}
{Heymans} C., {Van Waerbeke} L., {Miller} L., {Erben} T., {Hildebrandt} H.,
  {Hoekstra} H., {Kitching} T.~D., {Mellier} Y., {Simon} P., {Bonnett} C.,
  {Coupon} J., {Fu} L., {Harnois D{\'e}raps} J., {Hudson} M.~J., {Kilbinger}
  M., {Kuijken} K., {Rowe} B., {Schrabback} T., {Semboloni} E., {van Uitert}
  E., {Vafaei} S., {Velander} M., 2012, \mnras, 427, 146

\bibitem[{{Hoekstra} {et~al.}(2006){Hoekstra}, {Mellier}, {van Waerbeke},
  {Semboloni}, {Fu}, {Hudson}, {Parker}, {Tereno}, \& {Benabed}}]{hoekstra06}
{Hoekstra} H., {Mellier} Y., {van Waerbeke} L., {Semboloni} E., {Fu} L.,
  {Hudson} M.~J., {Parker} L.~C., {Tereno} I., {Benabed} K., 2006, \apj, 647,
  116

\bibitem[{{Kaiser} \& {Squires}(1993)}]{kaiser93}
{Kaiser} N., {Squires} G., 1993, \apj, 404, 441

\bibitem[{{Kaiser} {et~al.}(1995){Kaiser}, {Squires}, \&
  {Broadhurst}}]{kaiser95}
{Kaiser} N., {Squires} G., {Broadhurst} T., 1995, \apj, 449, 460

\bibitem[{{Kaiser} {et~al.}(2000){Kaiser}, {Wilson}, \& {Luppino}}]{kaiser00}
{Kaiser} N., {Wilson} G., {Luppino} G.~A., 2000, ArXiv Astrophysics e-prints

\bibitem[{{Kilbinger} {et~al.}(2013){Kilbinger}, {Fu}, {Heymans}, {Simpson},
  {Benjamin}, {Erben}, {Harnois-D{\'e}raps}, {Hoekstra}, {Hildebrandt},
  {Kitching}, {Mellier}, {Miller}, {Van Waerbeke}, {Benabed}, {Bonnett},
  {Coupon}, {Hudson}, {Kuijken}, {Rowe}, {Schrabback}, {Semboloni}, {Vafaei},
  \& {Velander}}]{kilbinger13}
{Kilbinger} M., {Fu} L., {Heymans} C., {Simpson} F., {Benjamin} J., {Erben} T.,
  {Harnois-D{\'e}raps} J., {Hoekstra} H., {Hildebrandt} H., {Kitching} T.~D.,
  {Mellier} Y., {Miller} L., {Van Waerbeke} L., {Benabed} K., {Bonnett} C.,
  {Coupon} J., {Hudson} M.~J., {Kuijken} K., {Rowe} B., {Schrabback} T.,
  {Semboloni} E., {Vafaei} S., {Velander} M., 2013, \mnras, 430, 2200

\bibitem[{{Kitching} {et~al.}(2008){Kitching}, {Miller}, {Heymans}, {van
  Waerbeke}, \& {Heavens}}]{kitching08}
{Kitching} T.~D., {Miller} L., {Heymans} C.~E., {van Waerbeke} L., {Heavens}
  A.~F., 2008, \mnras, 390, 149

\bibitem[{{Kochanek}(1990)}]{kochanek90}
{Kochanek} C.~S., 1990, \mnras, 247, 135

\bibitem[{{Miller} {et~al.}(2013){Miller}, {Heymans}, {Kitching}, {van
  Waerbeke}, {Erben}, {Hildebrandt}, {Hoekstra}, {Mellier}, {Rowe}, {Coupon},
  {Dietrich}, {Fu}, {Harnois-D{\'e}raps}, {Hudson}, {Kilbinger}, {Kuijken},
  {Schrabback}, {Semboloni}, {Vafaei}, \& {Velander}}]{miller13}
{Miller} L., {Heymans} C., {Kitching} T.~D., {van Waerbeke} L., {Erben} T.,
  {Hildebrandt} H., {Hoekstra} H., {Mellier} Y., {Rowe} B.~T.~P., {Coupon} J.,
  {Dietrich} J.~P., {Fu} L., {Harnois-D{\'e}raps} J., {Hudson} M.~J.,
  {Kilbinger} M., {Kuijken} K., {Schrabback} T., {Semboloni} E., {Vafaei} S.,
  {Velander} M., 2013, \mnras, 429, 2858

\bibitem[{{Miller} {et~al.}(2007){Miller}, {Kitching}, {Heymans}, {Heavens}, \&
  {van Waerbeke}}]{miller07}
{Miller} L., {Kitching} T.~D., {Heymans} C., {Heavens} A.~F., {van Waerbeke}
  L., 2007, \mnras, 382, 315

\bibitem[{{Peacock} {et~al.}(2006){Peacock}, {Schneider}, {Efstathiou},
  {Ellis}, {Leibundgut}, {Lilly}, \& {Mellier}}]{peacock06}
{Peacock} J.~A., {Schneider} P., {Efstathiou} G., {Ellis} J.~R., {Leibundgut}
  B., {Lilly} S.~J., {Mellier} Y., 2006, {ESA-ESO Working Group on
  ''Fundamental Cosmology''}. Tech. rep.

\bibitem[{{Press} {et~al.}(1992){Press}, {Teukolsky}, {Vetterling}, \&
  {Flannery}}]{Press.1992}
{Press} W.~H., {Teukolsky} S.~A., {Vetterling} W.~T., {Flannery} B.~P., 1992,
  Numerical recipes in c: The art of scientific computing. second edition

\bibitem[{{Sargent} {et~al.}(2007){Sargent}, {Carollo}, {Lilly}, {Scarlata},
  {Feldmann}, {Kampczyk}, {Koekemoer}, {Scoville}, {Kneib}, {Leauthaud},
  {Massey}, {Rhodes}, {Tasca}, {Capak}, {McCracken}, {Porciani}, {Renzini},
  {Taniguchi}, {Thompson}, \& {Sheth}}]{sargent07}
{Sargent} M.~T., {Carollo} C.~M., {Lilly} S.~J., {Scarlata} C., {Feldmann} R.,
  {Kampczyk} P., {Koekemoer} A.~M., {Scoville} N., {Kneib} J.-P., {Leauthaud}
  A., {Massey} R., {Rhodes} J., {Tasca} L.~A.~M., {Capak} P., {McCracken}
  H.~J., {Porciani} C., {Renzini} A., {Taniguchi} Y., {Thompson} D.~J., {Sheth}
  K., 2007, \apjs, 172, 434

\bibitem[{{Schneider} \& {Seitz}(1995)}]{schneider95}
{Schneider} P., {Seitz} C., 1995, \actaa, 294, 411

\bibitem[{{Schrabback} {et~al.}(2010){Schrabback}, {Hartlap}, {Joachimi},
  {Kilbinger}, {Simon}, {Benabed}, {Brada{\v c}}, {Eifler}, {Erben},
  {Fassnacht}, {High}, {Hilbert}, {Hildebrandt}, {Hoekstra}, {Kuijken},
  {Marshall}, {Mellier}, {Morganson}, {Schneider}, {Semboloni}, {van Waerbeke},
  \& {Velander}}]{schrabback10}
{Schrabback} T., {Hartlap} J., {Joachimi} B., {Kilbinger} M., {Simon} P.,
  {Benabed} K., {Brada{\v c}} M., {Eifler} T., {Erben} T., {Fassnacht} C.~D.,
  {High} F.~W., {Hilbert} S., {Hildebrandt} H., {Hoekstra} H., {Kuijken} K.,
  {Marshall} P.~J., {Mellier} Y., {Morganson} E., {Schneider} P., {Semboloni}
  E., {van Waerbeke} L., {Velander} M., 2010, \aap, 516, A63

\bibitem[{{Van Waerbeke} {et~al.}(2013){Van Waerbeke}, {Benjamin}, {Erben},
  {Heymans}, {Hildebrandt}, {Hoekstra}, {Kitching}, {Mellier}, {Miller},
  {Coupon}, {Harnois-D{\'e}raps}, {Fu}, {Hudson}, {Kilbinger}, {Kuijken},
  {Rowe}, {Schrabback}, {Semboloni}, {Vafaei}, {van Uitert}, \&
  {Velander}}]{waerbeke13}
{Van Waerbeke} L., {Benjamin} J., {Erben} T., {Heymans} C., {Hildebrandt} H.,
  {Hoekstra} H., {Kitching} T.~D., {Mellier} Y., {Miller} L., {Coupon} J.,
  {Harnois-D{\'e}raps} J., {Fu} L., {Hudson} M., {Kilbinger} M., {Kuijken} K.,
  {Rowe} B., {Schrabback} T., {Semboloni} E., {Vafaei} S., {van Uitert} E.,
  {Velander} M., 2013, \mnras

\bibitem[{{Van Waerbeke} {et~al.}(2000){Van Waerbeke}, {Mellier}, {Erben},
  {Cuillandre}, {Bernardeau}, {Maoli}, {Bertin}, {McCracken}, {Le F{\`e}vre},
  {Fort}, {Dantel-Fort}, {Jain}, \& {Schneider}}]{vanwaerbeke00}
{Van Waerbeke} L., {Mellier} Y., {Erben} T., {Cuillandre} J.~C., {Bernardeau}
  F., {Maoli} R., {Bertin} E., {McCracken} H.~J., {Le F{\`e}vre} O., {Fort} B.,
  {Dantel-Fort} M., {Jain} B., {Schneider} P., 2000, \aap, 358, 30

\bibitem[{{Viola} {et~al.}(2013){Viola}, {Kitching}, \& {Joachimi}}]{viola13}
{Viola} M., {Kitching} T., {Joachimi} B., 2013, ArXiv e-prints

\bibitem[{{Viola} {et~al.}(2011){Viola}, {Melchior}, \& {Bartelmann}}]{viola11}
{Viola} M., {Melchior} P., {Bartelmann} M., 2011, \mnras, 410, 2156

\bibitem[{{Walsh} {et~al.}(1979){Walsh}, {Carswell}, \& {Weymann}}]{walsh79}
{Walsh} D., {Carswell} R.~F., {Weymann} R.~J., 1979, \nat, 279, 381

\bibitem[{{White}(2005)}]{white05}
{White} M., 2005, Astroparticle Physics, 23, 349

\bibitem[{{Wittman} {et~al.}(2000){Wittman}, {Tyson}, {Kirkman},
  {Dell'Antonio}, \& {Bernstein}}]{wittman00}
{Wittman} D.~M., {Tyson} J.~A., {Kirkman} D., {Dell'Antonio} I., {Bernstein}
  G., 2000, \nat, 405, 143

\end{thebibliography}

\label{lastpage}

\end{document}